\documentclass[10pt,conference]{IEEEtran}

\usepackage{cite}
\usepackage{balance}
\usepackage{newtxtext,newtxmath}
\usepackage{amsmath,amsfonts}
\usepackage{textcomp}
\usepackage{algorithm}
\usepackage[noEnd=true,rightComments=false]{algpseudocodex}
\usepackage{mathtools}
\usepackage{hyperref}
\usepackage{cleveref}
\usepackage{booktabs}
\usepackage{multirow}
\usepackage{graphicx}
\usepackage{float}
\usepackage{tikz}
\usepackage{siunitx}
\usepackage{enumitem}

\usepackage{xspace}
\newcommand{\Name}{\textsc{Arborist}\xspace}
\newcommand{\Accel}{\textsc{ArboristAccel}\xspace}

\DeclareRobustCommand*\circled[1]{\tikz[baseline=(char.base)]{
            \node[shape=circle,draw,inner sep=0.5pt] (char) {\small #1};}}

\algnewcommand\algorithmiccontinue{\textbf{continue}}
\algnewcommand\Continue{\State \algorithmiccontinue}

\hypersetup{
  hidelinks,
  pdftitle={Arborist: Algorithm-Hardware Co-Design for Fast and Efficient Motion Planning},
  pdfauthor={Yaotian Liu, Lingyi Huang, Zishen Wan, Bo Yuan, Cheng Tan, and Jeff (Jun) Zhang},
  pdfsubject={59th IEEE/ACM International Symposium on Microarchitecture (MICRO 2026)},
  pdfkeywords={motion planning, hardware acceleration, algorithm-hardware co-design, Fast Marching Tree}
}

\begin{document}

\title{Arborist: Algorithm-Hardware Co-Design for Fast and Efficient Motion Planning}

\author{
\IEEEauthorblockN{
Yaotian Liu\IEEEauthorrefmark{1},
Lingyi Huang\IEEEauthorrefmark{2},
Zishen Wan\IEEEauthorrefmark{3},
Bo Yuan\IEEEauthorrefmark{2},
Cheng Tan\IEEEauthorrefmark{1}\IEEEauthorrefmark{4}, and
Jeff (Jun) Zhang\IEEEauthorrefmark{1}}
\IEEEauthorblockA{\IEEEauthorrefmark{1}Arizona State University, Tempe, AZ, USA}
\IEEEauthorblockA{\IEEEauthorrefmark{2}Rutgers University, Piscataway, NJ, USA}
\IEEEauthorblockA{\IEEEauthorrefmark{3}Columbia University, New York, NY, USA}
\IEEEauthorblockA{\IEEEauthorrefmark{4}Google, Mountain View, CA, USA}
\IEEEauthorblockA{
\texttt{yaotian\_liu@asu.edu},
\texttt{lingyi.huang@rutgers.edu},
\texttt{zishen.wan@columbia.edu},\\
\texttt{bo.yuan@soe.rutgers.edu},
\texttt{chengtan@google.com},
\texttt{jeffzhang@asu.edu}}
}

\maketitle

%%%%%% -- PAPER CONTENT STARTS-- %%%%%%%%

\begin{abstract}
Real-time motion planning must execute under strict latency and energy constraints on resource-limited platforms. This paper presents \textsc{Arborist}, an algorithm–hardware co-design framework that accelerates Fast Marching Tree (FMT*) motion planning through synergistic algorithmic and architectural innovations. At the algorithm level, \textsc{Arborist} introduces safe multi-tree expansion with global competition management and cross-tree optimality checking to expose inter-tree parallelism while preserving path quality. At the architecture level, our proposed \textsc{ArboristAccel} integrates
an \textsc{Arborist} Toolbox that provides hardware support for inter-tree parallelism; a KD-tree Subsystem for high-throughput near-neighbor search;
a Parallel-Query Retrieval module 
that balances multi-bank memory accesses to alleviate system-level memory bandwidth bottlenecks; and a Look-ahead Planning Engine that exploits intra-tree parallelism while preserving path quality by in-order commit.
Synthesized in 16 nm and operating at 500 MHz, \textsc{ArboristAccel} sustains more than $10^{3}\times$ speedup over CPU across all workloads (peaking at $2.8\times 10^{3}\times$ on 2D-Maze), and delivers geomean improvements of $8.8\times$ speedup, $4.6\times$ area efficiency, and $3.1\times$ power efficiency over an FMT* ASIC baseline. The corresponding factors relative to a state-of-the-art RRT*-based
MOPED accelerator are $8.6\times$/$1.0\times$/$1.8\times$, at comparable path cost. To the best of our knowledge, \textsc{ArboristAccel} is the first dedicated hardware accelerator for FMT*-based motion planning.

\end{abstract}


\section{Introduction}
Motion planning is a critical workload in robotics~\cite{wang2021survey, zhou2022review}, autonomous systems~\cite{schwarting2018planning, teng2023motion}, embodied AI~\cite{kim2025supernova, wan2025reca, huang2025dadu, yu2020building}, humanoid control~\cite{de2024humanoid}, and manufacturing~\cite{Ghafarian2023Review}, enabling real-time decision making in high-dimensional and unknown environments.
Motion planning seeks to compute a feasible and efficient trajectory from a start configuration to a goal configuration while avoiding collisions with obstacles. High-quality, low-latency planning is essential for reliable robotic navigation and safe autonomous operation.

However, integrating state-of-the-art motion planning into real systems remains challenging.
First, achieving high success rates and path quality in high-dimensional configuration spaces requires exploring many sampled nodes, incurring substantial computational overhead.
Second, real-time planning must execute locally: although cloud offloading can accelerate computation~\cite{ieee2022robotchips}, unpredictable network latency makes it unsuitable for safety-critical applications such as self-driving vehicles or agile drones~\cite{yu2020building,levin2019realtime}.
Third, stringent device-level constraints (such as limited battery capacity and compact form factors) restrict on-device computational resources~\cite{murray2016robot}, further exacerbating the challenge.

Among state-of-the-art motion-planning solutions, sampling-based planners~\cite{sampling-based, PRM, LaValle1998RapidlyexploringRT, karaman2011sampling, janson2015fastmarchingtreefast} have emerged as promising candidates for real-time deployment because they explore complex, high-dimensional configuration spaces efficiently and include asymptotically optimal variants.
In particular, Fast Marching Tree (FMT*)-based motion planning~\cite{janson2015fastmarchingtreefast,BFMT,GMT,xu2020informed,wang2023informed} stands out because (1) its dynamic-programming wavefront expansion yields near-optimal path quality~\cite{sampling-based}, and (2) its lazy collision checking significantly reduces collision-test overhead.
Unfortunately, despite these algorithmic advantages, FMT*-powered motion planners still suffer from the high computational cost of searching the nearest neighbors in high-dimensional real-world environments~\cite{exenberger2025caravan}.

In parallel, hardware acceleration has emerged as a promising approach to real-time motion planning. For instance, RACOD~\cite{bakhshalipour2022racod} massively parallelizes collision checks and introduces speculative exploration mechanisms that effectively accelerate search-based motion planners such as A*~\cite{hart1968formal}. MOPED~\cite{huang2024moped} uses custom hardware with restructured neighbor-search and speculative pipelines for sampling-based Rapidly-exploring Random Trees (RRT*)~\cite{karaman2011sampling}, improving motion-planning throughput and energy efficiency in unknown environments.
Nevertheless, existing hardware solutions overlook the algorithmic advantages of FMT*-based motion planning, motivating approaches that can deliver high real-time performance while maintaining path quality.

This work presents \Name, a novel algorithm-hardware co-design framework for hardware-accelerated parallel FMT* execution. Importantly, \Accel is not hard-wired to a single planner: by separating a fixed execution substrate from a reconfigurable control layer, the same hardware retargets across the FMT*-family without redesigning the underlying datapath.
Across 2D/3D (city, maze, drone) and high-DoF (5D/7D arm) benchmarks, \Accel delivers geomean improvements of $8.8\times$ speedup, $4.6\times$ area efficiency, and $3.1\times$ power efficiency over an FMT* ASIC baseline. The corresponding factors relative to a state-of-the-art RRT*-based
MOPED accelerator~\cite{huang2024moped} are $8.6\times$/$1.0\times$/$1.8\times$, at comparable path cost.

Our contributions can be summarized as follows:
\begin{itemize}[leftmargin=1em]
    \item We present \Accel, the first dedicated hardware accelerator for FMT*-based motion planning, targeting the dominant bottlenecks in nearest-neighbor search, tree expansion, and speculative look-ahead planning.
    \item We develop the \Name algorithm, a multi-FMT* motion planner whose tree splitting, cross-tree overriding, and competition management expose inter-tree parallelism while preserving path optimality via an admissible cross-tree optimality check.
    \item We design the \Accel architecture: an \Name Toolbox for inter-tree control, a KD-tree Subsystem with an exactly balanced constructor for dynamic replanning, a Parallel-Query Retrieval engine balancing multi-bank memory accesses, and a Look-ahead Planning Engine for intra-tree speculation with in-order commit.
    \item \Accel  is a programmable substrate for the FMT* family: reconfiguring only the control layer adapts the datapath to FMT*, BFMT*, IAFMT*, and IABFMT*, achieving a $294.6\times$ geometric-mean CPU speedup. 
    The on-chip constructor adds $3.3–8.0\times$ end-to-end speedup over CPU-side tree reconstruction for the iterative IAFMT*/IABFMT* variants under dynamic replanning.
    %The on-chip constructor is critical to end-to-end acceleration: replacing it with CPU-side tree reconstruction limits speedup to $3.3–8.0\times$ across the four planners.
    \item Our evaluation shows that \Name preserves path quality and sustains more than $10^{3}\times$ CPU speedup across all workloads, with geomean $8.8\times$/$4.6\times$/$3.1\times$ speedup/area/power gains over an FMT* ASIC baseline. The corresponding factors relative to the RRT*-based
MOPED accelerator~\cite{huang2024moped} are $8.6\times$/$1.0\times$/$1.8\times$.
\end{itemize}

\section{Background and Motivation}
This section reviews motion-planning algorithms and prior hardware accelerators.

\subsection{Motion Planning}
Motion planning is the computational process of determining a robot’s trajectory from an initial to a goal configuration while avoiding obstacles in the environment. It involves computing a feasible, collision-free path that satisfies geometric, kinematic, and dynamic constraints~\cite{latombe2012robot}.
Classical formulations include geometric path planning (e.g., the “piano mover’s problem”)~\cite{SpatialPlanning}, kinodynamic planning that accounts for system dynamics and actuation limits~\cite{Kinodynamic}, and optimal motion planning, which seeks to minimize a cost functional such as path length or energy consumption~\cite{karaman2011sampling}.
Over the past four decades, the field has evolved through several major paradigms — from early combinatorial approaches with exponential-time guarantees~\cite{complexity_of_mp}, to potential field methods~\cite{1087247}, and, more recently, to sampling-based motion planning (SBMP) algorithms~\cite{sampling-based}, which scale more efficiently to high-dimensional configuration spaces.

SBMP algorithms such as Probabilistic Roadmaps (PRM)~\cite{PRM} and Rapidly-exploring Random Trees (RRT)~\cite{LaValle1998RapidlyexploringRT} generate feasible solutions by incrementally exploring the configuration space through random sampling. These methods provide \emph{probabilistic completeness}, ensuring eventual success if a feasible path does exist. Subsequent advancements introduced asymptotically optimal variants such as PRM*~\cite{karaman2011sampling} and RRT*~\cite{karaman2011sampling}, which converge toward cost-optimal solutions as the number of samples increases.
Building on this line of work, the Fast Marching Tree (FMT*) algorithm~\cite{janson2015fastmarchingtreefast} integrates dynamic programming principles with sampling-based search to achieve faster convergence and fewer collision checks, offering improved scalability in complex environments.
Today, these algorithms underpin a wide range of applications, including robotic manipulation~\cite{5152399}, autonomous driving~\cite{5175292}, and multi-robot coordination~\cite{6630671}, where efficient and reliable planning is essential for real-time operation.

% \begin{figure}[ht]
%   \centering
%   \begin{subfigure}{\columnwidth}
%     \includegraphics[width=\textwidth]{figs/comparison_curves.pdf}
%     \caption{Sampling based method comparison. Shadowed area for min/max range indication. Percentage annotation is for the Success Rate, if not 100\%.
%     %\jeff{@lingyi: with this data on RRT, we should be able to get the scaling factor from MOPEd and add comparison in our evaluation? } 
%     \yt{ADD information about different methods, why compared to those? OMPL, use Apartment only, how to calculate success rate. which CPU. etc.}}
%     \label{fig:alg_comparisn}
%   \end{subfigure}

%   \vspace{5pt} % spacing between the two figures

%   \begin{subfigure}{0.95\columnwidth}
%     \includegraphics[width=\textwidth]{figs/profiling_pie_charts.pdf}
%     \caption{Distribution comparison. FMT* is different compared to RRT*, with Neighbor Search takes up most of the time. \jeff{@Yaotian, I need a bit more info: how many samples are used for each, and what platform do we profile this?} \yt{match with MOPED, and reference to the github}}
%     \label{fig:pie}
%   \end{subfigure}

%   \caption{Comparison between sampling methods }
%   \label{fig:alg_comparison_combined}
% \end{figure}

\subsection{Fast Marching Tree (FMT*) Family}
\label{sec:bg-fmt-family}

\begin{figure}[t] 
\centerline{\includegraphics[width=\columnwidth]{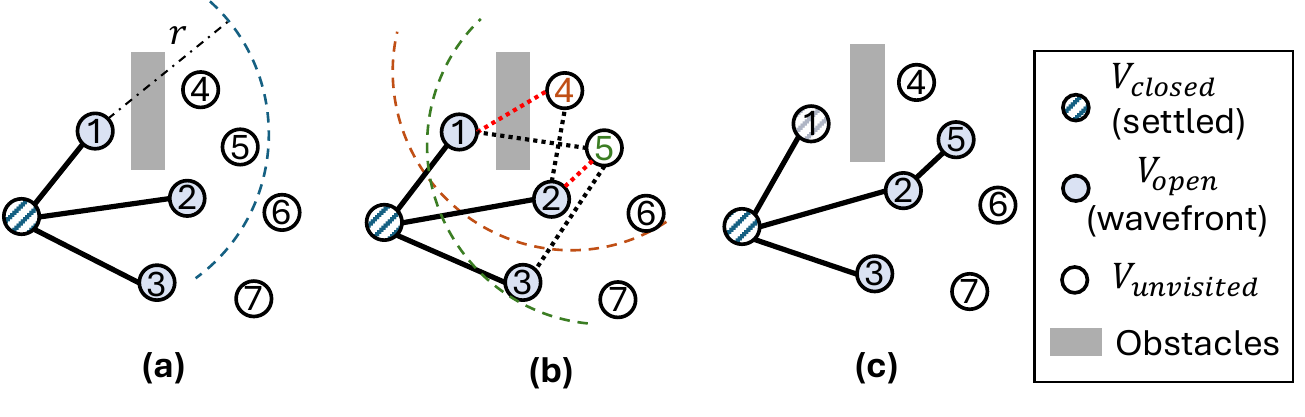}}
    \caption{
    %\yt{Naming (a, b, c), (a) is sequential originally}
    %\jeff{@Yaotian: let's discussion about the naming of Vopen?} \yt{Not yet, but we should though...}
    Example of one FMT* iteration. (a) Select the \textit{minimum-cost} node in $V_{{open}}$ (i.e., node \circled{1}) and identify its unvisited neighbors within radius $r$ (nodes \circled{4} and \circled{5}). (b) For each unvisited neighbor, consider candidate parents in $V_{open}$ within radius $r$ (dashed lines), and minimize the parent's cost-to-come plus the connecting edge cost (\circled{4} $\to$ \circled{1} and \circled{5} $\to$ \circled{2}, highlighted by red dashed lines). (c) Perform collision checking only for these selected connections and add the valid edges (only \circled{5} $\to$ \circled{2} succeeds). Finally, move \circled{1} to $V_{{closed}}$ and insert \circled{5} into $V_{{open}}$.
    For clarity, we will refer to $V_{\text{closed}}$ as ``settled points'' and $V_{\text{open}}$ as ``wavefront points'' in the rest of the paper.
    }
    % \vspace{-15pt}
    \label{fig:fmt}
\end{figure}

% The Fast Marching Tree (FMT*) algorithm~\cite{janson2015fastmarchingtreefast} is a recent addition to the family of asymptotically optimal motion planners. 
% Unlike RRT*, which incrementally grows a tree through random sampling, FMT* performs a forward dynamic programming recursion over a pre-sampled set of nodes. It expands a ``wavefront'' of explored states—analogous to numerical fast marching methods in continuous domains~\cite{Sethian1996AFM}—adapted for discrete, sampling-based planning. \jeff{@Yaotian: do we have a time complexity of FMT compare with RRT?}

The Fast Marching Tree (FMT*) algorithm~\cite{janson2015fastmarchingtreefast} grows a single tree via a lazy dynamic-programming recursion over an implicit random geometric graph, expanding outward in cost-to-come ``wavefront'' (akin to Fast Marching) while constructing and searching the graph simultaneously. It is asymptotically optimal under standard robustness assumptions with a proven convergence-rate bound, and runs in expected $O(n\log n)$ time with $O(n)$ collision checks.

%\subsection{Derivatives and Extension of FMT*}
 
Since its introduction, FMT* has inspired a broad range of derivative algorithms that extend its wavefront expansion principle to various settings.
Bidirectional FMT* (BFMT*)~\cite{BFMT} grows two trees simultaneously from the start and goal to accelerate convergence.
Group Marching Tree (GMT*)~\cite{GMT} extends FMT* to parallel architectures by employing group updates and approximate dynamic programming, enabling efficient execution on GPUs.
Informed Anytime Fast Marching Tree (IAFMT*)~\cite{xu2020informed} augments FMT* with an anytime mechanism that alternates incremental search and dynamic optimal search to keep improving a found path. Kino-FMT*~\cite{allen2015toward} applies FMT* to generic kinodynamic problems by handling directed connections and two-point BVP/steering.
Many of these FMT*-based planners have been integrated into the Open Motion Planning Library (OMPL)~\cite{sucan2012the-open-motion-planning-library} as standard planning methods, underscoring the prominence of FMT* as a foundational algorithm in modern motion planning.
% Heuristic-guided variants such as Batch Informed Trees (BIT*)~\cite{BIT} and Advanced BIT* (ABIT*)~\cite{ABIT} integrate FMT*-like expansion with cost-to-go heuristics for more aggressive pruning of the search space.
% More recent algorithms, including AIT~\cite{Strub2020AdaptivelyIT}, EIT~\cite{EIT}, and others~\cite{FIT, BiAIT, Chen2021GreedyB}, further advance this lineage.

\noindent\textbf{Why FMT*?}
Figure~\ref{fig:fmt} illustrates one iteration of FMT*. It offers two key advantages:
(i) it reduces reliance on costly collision checks by adopting lazy edge validation, deferring collision evaluation until necessary, and
(ii) each minimum-cost frontier expansion can connect multiple unvisited neighbors.
As a result, FMT* achieves faster convergence toward high-quality solutions than RRT* and PRM*, particularly in high-dimensional or cluttered environments.
Empirical evaluations~\cite{sampling-based} demonstrate that FMT* consistently delivers competitive success rates and lower path costs across diverse planning benchmarks.

\begin{figure}[t]  
\centerline{\includegraphics[width=\columnwidth]{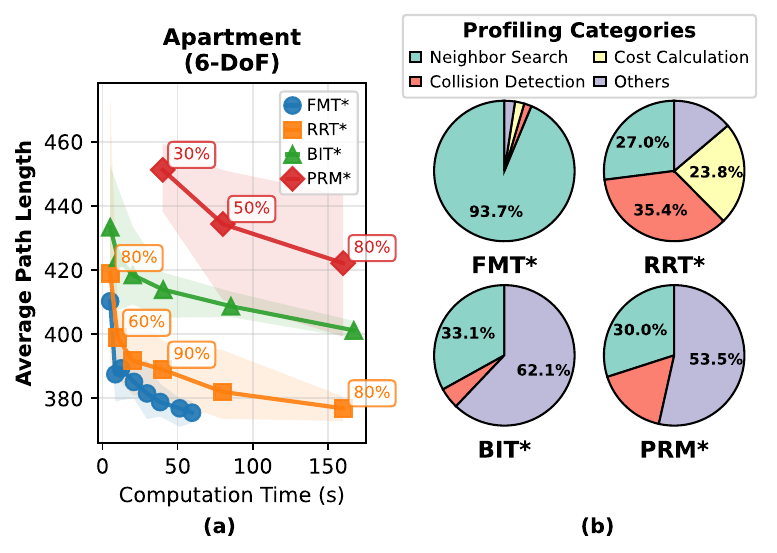}}
    \caption{
    (a) Comparison of sampling-based motion planners. Shaded regions indicate minimum–maximum ranges; percentages indicate success rates below 100\%.
    (b) Runtime breakdown. Unlike RRT*, FMT* spends most of its runtime on neighbor search.
    We select PRM*~\cite{karaman2011sampling} as a classic baseline, RRT*~\cite{karaman2011sampling} as a ubiquitous single-query tree with rewiring, BIT*~\cite{BIT} as the seminal informed-tree planner, and FMT*~\cite{janson2015fastmarchingtreefast} as the seminal wavefront dynamic-programming planner. Success rates are measured over 10 runs. FMT* uses 5k–30k samples; the other planners use time limits of 5–160\,s. Experiments use OMPL~\cite{sucan2012the-open-motion-planning-library} on an AMD Ryzen Threadripper 3960X CPU at 3.8\,GHz.
    }
    \label{fig:comb_profiling}
\end{figure}

%\jeff{@01, can you help write a something to describe Figure~\ref{fig:comb_profiling} as we discussed over phone? We are using this fig to motivate our work} 
%\jeff{@yaotian, check 01 writting below}\yt{checked, all good}
%Figure~\ref{fig:comb_profiling} further supports this choice with empirical data.
Our profiling in Figure~\ref{fig:comb_profiling}(a) shows that FMT* achieves the shortest average path length with the lowest computation time among alternative sampling-based planners, confirming that it offers an excellent trade-off between path quality and runtime.
Figure~\ref{fig:comb_profiling}(b) breaks down the runtime and shows that FMT* spends only a small fraction of its work on collision checking, whereas other planners are dominated by this cost.
Since collision checking is well known to be the primary bottleneck in sampling-based motion planning \cite{bakhshalipour2022racod,murray2019programmable,lian2018dadu,li2021anytime}, an algorithm that structurally minimizes collision checks is particularly attractive for on-device accelerators.
Thus, FMT* not only provides strong algorithmic performance, but also a more hardware-friendly cost profile for our accelerator design.

%\jeff{@01, help clean up the following with Yaotian's points above. Now we want to motivate why we are doing our \Name }

%However, there are still bottleneck for FMT* from fig. 1 --> emphasize sequential execution, in-regular memory access, branches, etc? 
%--> emphasize which are common in all sampling based approaches. 
%The original FMT*, however, is inherently sequential due to the recursive tree traversal \yt{due to its Dynamic Programming}. 

%\subsection{Why FMT*?}

%\subsection{Hardware Acceleration for SBMP}
\subsection{Motivation for Hardware-accelerated Motion Planning}

\noindent\textbf{Real-time, Power, and End-to-End Latency Requirements.}
Real-time responsiveness at sub-second, often millisecond, scales is essential for safe deployment in safety-critical environments~\cite{zhou2020robust}: industrial robots must react to obstacles or human motion within tightly bounded windows, making \emph{worst-case} timing a first-class constraint. Concretely, agile aerial robots run MPCC controllers at 100\,Hz with online replanning every step~\cite{romero2022replanning}, and topological replanning is constrained to 5--7\,ms of search plus 10\,ms of optimization~\cite{zhou2020robust}. At the same time, battery-powered platforms such as nano-drones (10\,cm, 27\,g)~\cite{palossi2019autonomous} impose stringent power and form-factor budgets. Because total response latency composes perception, planning, and control, delays in \emph{any} stage propagate through the closed loop and degrade overall responsiveness~\cite{pendleton2017perception}, an especially dangerous failure mode in safety-critical scenarios.

\noindent\textbf{Motion Planning is the Computational Bottleneck.}
With perception and control already heavily optimized~\cite{kim2025supernova,huang2024algorithm}, motion planning has emerged as the dominant bottleneck of perception--planning--control pipelines~\cite{wan2025reca,huang2025dadu}. As shown in Figure~\ref{fig:comb_profiling}, even optimized planners take tens of seconds on 6-DoF scenes; GPU implementations still need tens of seconds for a Kuka 7-DoF arm~\cite{yu2021reducing}; on a NVIDIA Jetson TX-2, planning runs \textbf{$3.5$--$8\times$} slower than perception+control on MAVBench workloads~\cite{boroujerdian2018mavbench}; and ARM Cortex-class CPUs spend \textbf{${\sim}5$\,s} and \textbf{${\sim}1$\,W} per indoor-navigation query, while typical mobile micro-robots demand \textbf{$10\times$} lower power and \textbf{$100\times$} higher throughput~\cite{8004525}. These large gaps motivate specialized hardware for motion planning.

\noindent\textbf{Hardware Acceleration Challenges.}
From a computational perspective, SBMP (such as FMT*) is fundamentally different from conventional AI or vision workloads: it features irregular, data-dependent computation with poor spatial/temporal locality and high control-flow divergence, while runtime is dominated by latency-variable nearest-neighbor and collision queries over large spatial structures. Moreover, its parallelism is largely unstructured because the active node set evolves dynamically, preventing efficient load balancing on GPUs.

\begin{figure}[t]
\centerline{\includegraphics[width=\columnwidth]{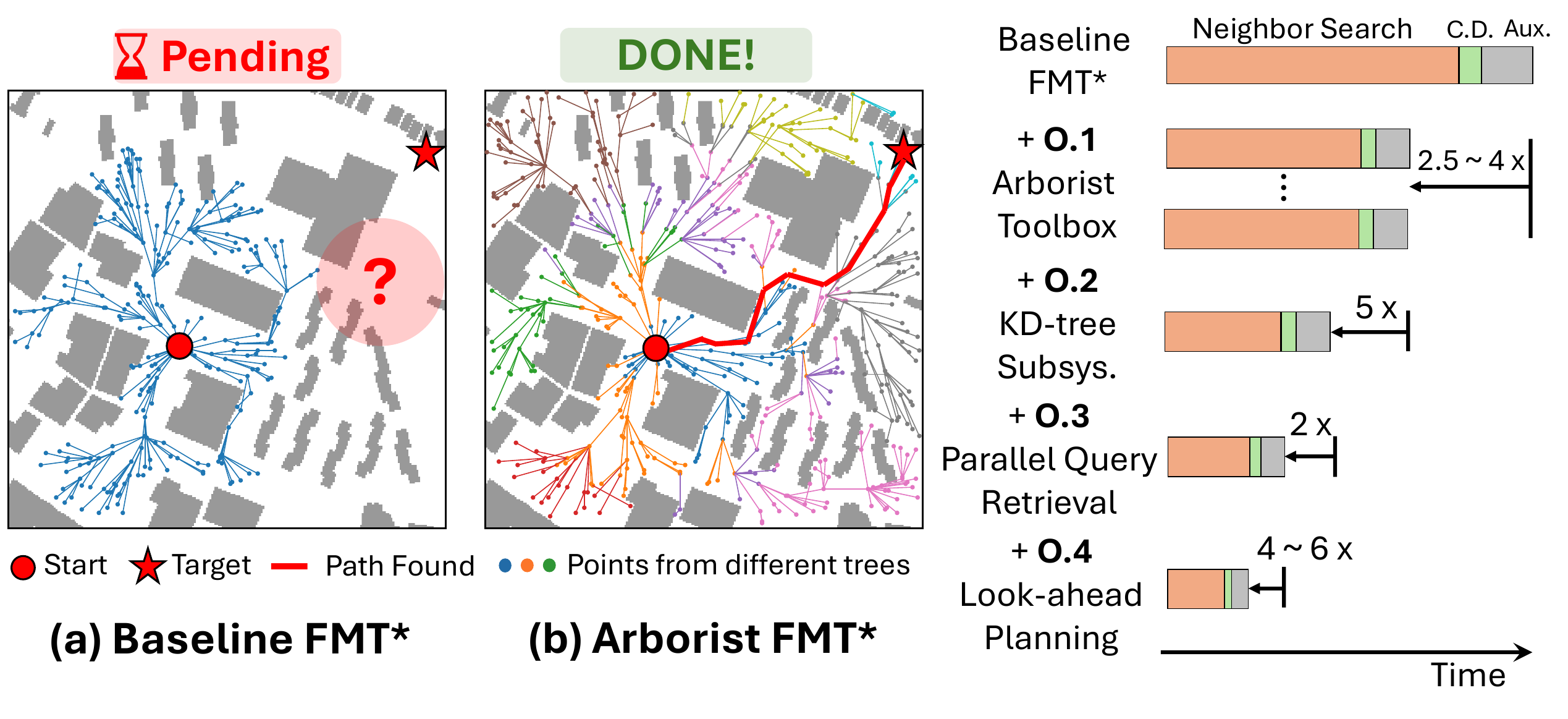}}
    \caption{\Name vs. baseline FMT*. Baseline FMT* is inherently sequential due to its dynamic-programming wavefront recursion, leaving compute resources underutilized. \Name unlocks both \emph{inter-tree} parallelism (via safe tree-splitting with global competition management) and \emph{intra-tree} parallelism (via look-ahead planning), achieving substantially faster computation at identical path quality.}
    \label{fig:arborist-example}
\end{figure}

Together, these characteristics render motion planning a highly irregular, memory-intensive, and latency-sensitive workload, demanding novel architectural mechanisms that can efficiently handle data-dependent computation, dynamic memory access, and real-time responsiveness. This paper presents \Name, an algorithm-hardware co-design framework integrated with a set of novel optimizations (as shown in~\Cref{fig:arborist-example}) for fast and efficient FMT* motion planning accelerator design.

%\jeff{@yaotian, check 01 writing for below} \yt{checked}
%\jeff{check}

% \noindent\textbf{Existing Solutions.}
% A number of prior works have proposed specialized accelerators for motion planning and its key subroutines.
% For end-to-end motion-planning accelerators, \cite{murray2016microarchitecture} presents one of the first architectures targeting PRM~\cite{PRM}, while MOPED~\cite{huang2024moped} co-designs algorithm and hardware for RRT*~\cite{karaman2011sampling} to reduce both the frequency and the cost of collision checks.
% RACOD~\cite{bakhshalipour2022racod} couples a collision-detection accelerator with run-ahead search to significantly speed up grid-based A* planning~\cite{hart1968formal}.
% MPAccel~\cite{shah2023energy} focuses on accelerating collision checking, using spatially aware scheduling with early-exit collision units to eliminate redundant work.
% In parallel, KD-tree–based neighbor search has also been widely accelerated: Tigris~\cite{xu2019tigris} and QuickNN~\cite{pinkham_quicknn_2020} both exploit query-level and node-level parallelism to speed up KD-tree traversal.

%\begin{figure}[htbp]  \centerline{\includegraphics[width=0.9\columnwidth]{figs/multi-example.pdf}}
%\caption{\Name Example ~\jeff{jeff to think if move this figure to the first page?}}
%\label{fig:arborist-example}
%\end{figure}

% \section{Methods}

% \begin{figure*}[htbp]
%   \centerline{\includegraphics[width=\textwidth]{figs/multitree.pdf}}
%     \caption{\Name~~Framework (+ Termination logic sketch)}
%     \label{fig:multi-tree-management}
% \end{figure*}

\section{\Name Algorithm}
\label{sec:arborist}

%  \begin{figure*}[t]
%   \centering
%   \begin{subfigure}[t]{0.86\textwidth}
%     \centering      \includegraphics[width=\linewidth]{figs/multitree_25Nov_V2.pdf}
%     \caption{}
%     \label{fig:arborist-skip-management}
%   \end{subfigure}\hfill
%   \begin{subfigure}[t]{0.14\textwidth}
%     \centering
% \includegraphics[width=\linewidth]{figs/preview-crosstree-check.pdf}
%     \caption{}
%     \label{fig:pre-crosstree-check}
%   \end{subfigure}
%   \caption{\Name Algorithm Framework.}
%   \label{fig:arborist}
% \end{figure*}

\begin{figure*}[t]
    \centerline{\includegraphics[width=\textwidth]{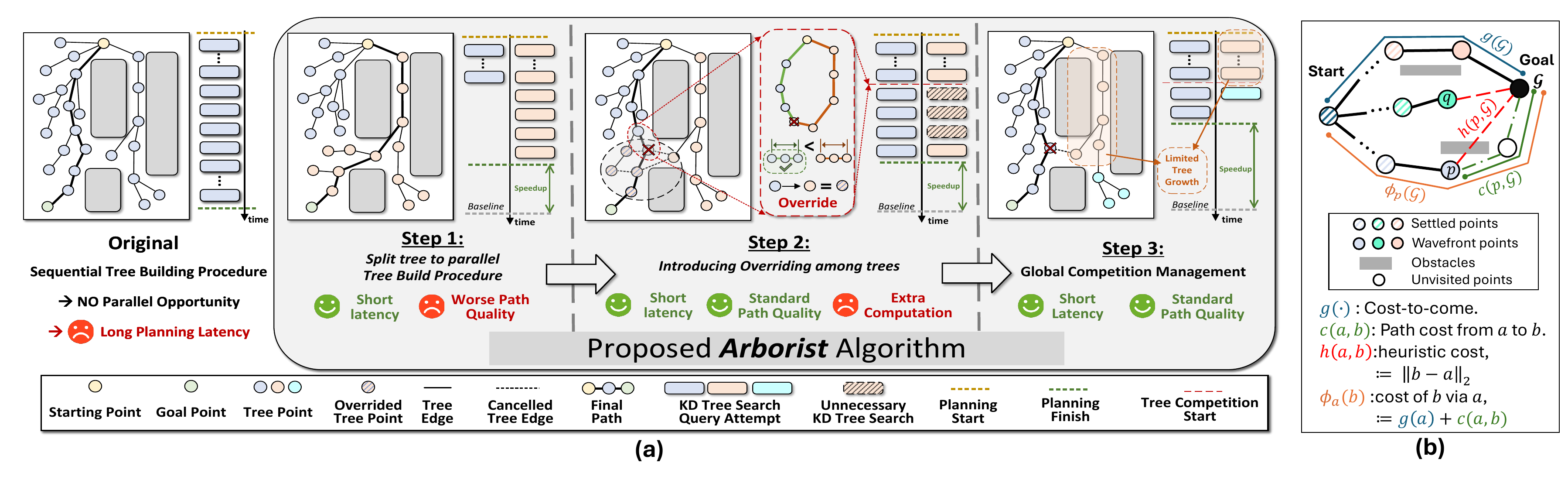}}
    \caption{(a) Tree splitting, overriding, and global competition management enable parallel multi-tree execution while preserving path quality. (b) Cross-tree optimality check prunes trees that cannot improve the current goal cost, enabling correct termination.}
    \label{fig:arborist}
\end{figure*}

\begin{figure}[t]
    \centerline{\includegraphics[width=\columnwidth]{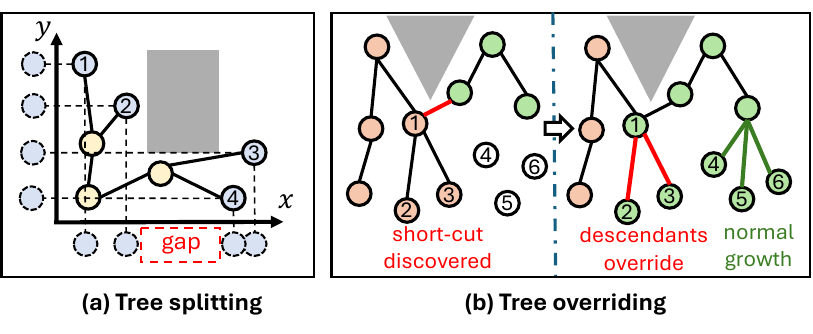}}
    \caption{(a) Tree-splitting detection by finding the gap along each axis. (b) Tree overriding: when a shortcut is discovered, its descendants are overridden in the same process of tree growth, alongside normal growth.}
    \label{fig:split-override}
\end{figure}

This section introduces the \Name~algorithm framework, which exploits environmental context to unlock novel parallelism opportunities in the otherwise sequential FMT* algorithm without sacrificing path quality. As illustrated in~\Cref{fig:arborist-example}, baseline FMT* is inherently sequential due to its recursive tree traversal, while \Name enables both inter-tree (safe tree-splitting) and intra-tree (look-ahead planning) parallelism, yielding faster computation at identical path quality.

An overview of the \Name~algorithm framework is shown in \Cref{fig:arborist}; the four steps are detailed in the following subsections.

%we propose a Cross-tree Optimality Check ,  performs when one tree hits the target, to ensure path optimality while maximizing efficiency. 

% obstacles in real-world maps can divert the growth of FMT. In many cases, these diverted branches evolve independently without interacting with others, creating opportunities for parallel expansion. To exploit this, we propose Multi-FMT, which breaks the inherently sequential nature of FMT by identifying and executing such parallel opportunities.

% Multi-FMT consists of three key components:

% \begin{enumerate}
% \item \textbf{Tree Split}: determining when to split a tree so that multiple branches can grow simultaneously.
% \item \textbf{Tree Management}: coordinating and merging trees when they converge.
% \item \textbf{Termination}: detecting when the optimal path has been found and terminating the search.
% \end{enumerate}

\subsection{Step 1: Tree Splitting}
\label{subsec:tree-split}
\noindent\textbf{Intuition.} Sampling nodes in the FMT* propagating wavefront (i.e., $V_{\rm open}$ in \Cref{fig:fmt}) are naturally partitioned into clusters by environmental contexts such as obstacles. Arborist leverages this property to enable opportunities for simultaneous multi-tree expansion (i.e., inter-tree parallelism).

As shown in \Cref{fig:arborist} (a) (Step 1), \Name exploits this clustering behavior in $V_{\rm open}$ to enable parallel multi-tree growth. A straightforward approach would construct the minimum spanning tree (MST)~\cite{enwiki:1317796852} of sampled nodes and detect large edge outliers as natural cluster boundaries. However, this method incurs $O(n^2D)$ complexity for pairwise distance computation and an additional $O(n^2)$ for MST construction, making it unsuitable for real-time planning. Therefore, in our approach, for each state dimension $d$, it projects nodes onto coordinate $x[d]$, sorts the projected values ($O(n \log n)$), and computes the largest adjacent projected-coordinate gap $G_d$ ($O(n)$). This gap is purely geometric and is not based on path cost or heuristic values. As shown in \Cref{fig:split-override} (a), we project the current wavefront onto $x$ and $y$. While the $y$ projection is evenly distributed, the $x$ projection shows a clear gap, indicating a possible splitting opportunity. If $\max_d G_d \ge \beta r_n$, \Name partitions the tree at the midpoint of the largest qualifying gap, where $r_n$ is the FMT* neighborhood radius and $\beta$ controls split aggressiveness. The split enables independent sub-tree expansion.
% for each dimension, \Name sorts nodes ($O(n \log n)$) and identifies the largest inter-node gap ($O(n)$). If a sufficiently large gap exists along any dimension, \Name partitions the node set accordingly, enabling independent sub-tree expansion.
This approach provides a simple yet efficient approximation of high-dimensional clustering, avoiding the heavy cost of MST computation. 

% \begin{hlblock}
% However, tree splitting must evaluate a per-axis histogram to find empty intervals (“gaps”),
% making its cost proportional to the dimensionality $D$.
% To reduce this overhead, \Name ranks axes using a lightweight \emph{variance-skew} metric
% derived from a Principal Component Analysis (PCA)~\cite{PCA_Wikipedia} on the sample points.
% Let $v_d$ denote the fraction of total sample variance explained by axis $d$ (its variance ratio).
% For an axis-balanced (isotropic) distribution, $v_d \approx 1/D$.
% We score each axis by $s_d = \left| v_d - \frac{1}{D} \right|$,
% and check axes in descending $s_d$, prioritizing directions with strong structure:
% either \emph{low variance} (often caused by obstacle-induced constraints) or \emph{high variance}
% (indicating an elongated extent).

% To quantify how much splitting signal is retained when checking only the top-$K$ axes,
% we define the per-axis \emph{gap magnitude} $g_d$ as the longest empty run in the axis histogram,
% and the \emph{gap coverage} of the top-$K$ axes as
% \begin{equation}\label{eq:gap_cov}
%   \mathrm{GapCov}(K) =
%   \frac{\sum_{d \in \mathrm{top}\text{-}K} g_d}{\sum_{d=1}^{D} g_d}.
% \end{equation}
% Thus, by checking only the top-$K$ axes, the hardware can skip the remaining $(D-K)$ histogram
% pipelines with minimal loss in gap coverage.

% \end{hlblock}

One potential issue with a naive, greedy tree-splitting strategy is that it might result in suboptimal path planning.
In FMT*, tree expansion rate is determined by the number of nodes updated per wavefront rather than the cumulative path cost. As a result, trees expand faster in sparse regions than dense ones. In \Cref{fig:arborist} (a) (Step 1), the orange branch in a narrow passage advances rapidly and consumes nearby samples before the blue branch arrives, preventing the latter—despite representing the optimal path—from reaching the goal.
This imbalance highlights a key limitation of naive splitting: while it improves parallelism, it can degrade solution quality due to uncoordinated competition among trees.
% Empirically, as shown in \Cref{fig:arborist-steps}, Step 1 reduces processing time by 60.4 \% but yields a path that is 27.4 \% longer than the optimal solution. 
%Therefore, a robust coordination mechanism is required to preserve path quality under concurrent expansion. 

%However, only splitting is not enough. In FMT*, expansion rate is governed by the number of points updated per wavefront, not path cost, so sparse regions advance faster than dense ones. In \Cref{fig:arborist} (a) (Step~1), the orange branch in a narrow passage grows quickly and consumes nearby samples before the blue branch arrives, starving it and reaching the goal first. The optimal path should come from the blue branch, but is precluded by this resource capture. Hence, naive splitting alone cannot preserve path quality; explicit competition management is required. In our experiment, as shown in \Cref{fig:arborist-steps}, Step~1 reduces processing time by 60.4\% but produces a path that is 27.4\% longer than the optimal solution. Therefore we require a robust mechanism to guarantee that solution quality is not compromised. To address this, we introduce Tree Competition.

\subsection{Step 2: Tree Overriding}
\label{subsec:tree-competition}

To mitigate the quality degradation from Step 1, \Name introduces Step 2, a tree-overriding mechanism that dynamically prioritizes trees based on path cost and allows a slower tree to reclaim nodes from a faster one when doing so improves the global solution.

When a tree attempts to connect to a node, it first checks whether that node has already been claimed by another tree. If so, the connection with the lower path cost ($g(\cdot)$) takes precedence, overriding the previous assignment. For example, in \Cref{fig:arborist} (a) (Step 2), the green path has a lower path cost than the red one, prompting the conflicting red node to be overridden. This tree overriding mechanism enforces global path optimality by ensuring that only the most cost-efficient connections persist.

\noindent\textbf{Handling Descendants of Overridden Nodes.} When an override occurs, we do not immediately update its descendants. Instead, propagation updates nodes as the wavefront passes through them again; if a better cost is found, the node is updated. This approach requires each node to have only a single parent pointer to maintain the tree structure, rather than multiple pointers to all its descendants. As shown in \Cref{fig:split-override} (b), a shortcut from the green tree to the orange tree at point 1 is first discovered, triggering an override that assigns point 1 to the green tree. Subsequently, the green tree follows the growth procedure, which simultaneously performs descendant overrides (points 2 and 3) and normal growth (points 4, 5, and 6).

Therefore, tree overriding inevitably introduces additional computational overhead because the overriding wavefront passes through previously connected nodes again. Furthermore, suboptimal trees continue to grow, generating additional branches that must later be overridden and further increasing computational overhead.

\subsection{Step 3: Global Competition Management}
\label{subsec:competition-management}

When a connection override occurs from Step 2, the affected (victim) tree enters a \textit{hibernation} state. During hibernation, the tree 
suspends expansion but remains visible to other trees, allowing further overrides if a better connection emerges.
A hibernating tree is reactivated once its wavefront cost drops below that of the overriding tree, indicating that the competing tree can no longer yield a superior path.
Meanwhile, computational resources (e.g., for neighbor search and collision checking) from hibernating trees are dynamically reallocated to active ones.
This mechanism effectively eliminates unnecessary computation from Step 2 (tree overriding) and improves overall efficiency.

\noindent\textbf{Hibernation vs. Purging.}
Hibernation differs from \textit{purging} (permanent deactivation) in that it preserves potentially optimal subtree structures: even when a single suboptimal connection triggers tree-wide suspension, other portions may still be optimal and can later reactivate once the tree's wavefront cost becomes favorable. As shown in Step~3 of \Cref{fig:arborist} (a), the light-blue branch resumes growth precisely because it was hibernated rather than purged.

\noindent\textbf{Algorithm.}
\Name Competition Management is enabled by two functions:

\begin{itemize}[leftmargin=1em]\label{item}
\item \textit{PropagatingPoints($z$): find possible points to propagate.} A tree may grow to a node $n$ if $n\in V_{\rm unvisited}$ \emph{or} $n$ belongs to another tree. To eliminate futile competition, we propagate only when $g(n)>g(z)$, where $g(\cdot)$ is the cost, with $g(V_{\rm unvisited})= +\infty$.

\item \textit{ResolveClaim($x$): resolves ownership (claim vs. override) and triggers hibernation if needed.} Insert node $x$ normally if $x\in V_{\rm unvisited}$. Otherwise, if $x\in V_{\rm open}\cup V_{\rm closed}$, update $g(x)$ and its parent only when the new cost is lower; if $x$’s original tree remains active, place it in \emph{hibernation}.
%to halt non-optimal growth  and free up resources.
\end{itemize}

% \begin{algorithm}[htbp]
% \caption{Termination}\label{alg:termination}
% \begin{algorithmic}[1]
%     \State{}

%     \Function{Termination}{$z$}
%         \If{$z$ is $x_{goal}$}
%             \If {cost of goal $<$ costs of all active trees}
%                 \State{\Return \textbf{success}}
%             \Else 
%                 \LComment{Another tree may still discover a shorter path to the goal.}
%                 \State {reached $\gets$ true}
%                 \State {Purge all trees with cost higher than $z$'s cost, including current tree $\mathcal{T}$}
%                 \State \textbf{continue}
%             \EndIf
%         \EndIf
%         \If {reached}
%             % \LComment{Impossible to find a path with lower cost}
%             \State {\textbf{If} current cost of goal $<$ all active trees' costs, \Return \textbf{success}}
%         \EndIf
%     \EndFunction
% \end{algorithmic}
% \end{algorithm}

\subsection{Step 4: Cross-tree Optimality Check}
\label{subsec:cross-tree-check}

% \begin{figure}[t]
%     % \vspace{-5pt}
%     \centerline{\includegraphics[width=\columnwidth]{figs/termination.pdf}}
%     \caption{Cross-tree Optimality Check Explanation.}
%     \label{fig:cross-tree-check}
% \end{figure}

\noindent\textbf{Motivation.} Because tree propagation speed in FMT* is not directly correlated with path cost, the first tree to reach the goal may not produce the optimal solution. Simply adopting the original FMT* termination rule~\cite{janson2015fastmarchingtreefast}, i.e., stopping once the goal point is selected, often yields suboptimal paths.

%As discussed earlier, the propagation speed of a tree is not consistent with respect to cost. Therefore, the first tree to reach the goal does not necessarily yield the lowest-cost path. If we simply adopt the original FMT*~\cite{janson2015fastmarchingtreefast} termination rule, stopping as soon as the goal point is selected, the resulting path is often suboptimal.

% To address this, \Name introduces termination logic (TL) tailored for multi-tree execution, termed the Cross-tree Optimality Check (\Cref{fig:arborist}(b)). When the goal $\mathcal{G}$ is first reached with cost $g(\mathcal{G})$, \Name evaluates the wavefront nodes of every retained tree $\mathcal{T}$, including hibernating trees. A tree is pruned only when $g(x)+h(x,\mathcal{G})\ge g(\mathcal{G})$ for every node $x$ in its wavefront, equivalently when the minimum of these lower bounds cannot improve the incumbent goal cost. Here, $h(x,\mathcal{G})$ is an admissible lower bound on the remaining cost to the goal.
To address this, \Name introduces a termination logic (TL) tailored for multi-tree execution, termed the Cross-tree Optimality Check (\Cref{fig:arborist} (b)). When the goal $\mathcal{G}$ is first reached with cost $g(\mathcal{G})$, \Name evaluates every active tree $\mathcal{T}$ and its wavefront nodes $x$, checking whether $\phi_x(\mathcal{G}) \equiv g(x) + c(x, \mathcal{G}) \ge g(x) + h(x,\mathcal{G}) \ge g(\mathcal{G})$. If so, no path through $x$ can improve the current solution, since $h(x,\mathcal{G})$ is an admissible heuristic, a lower bound on the true cost $c(x,\mathcal{G})$. Thus, tree $\mathcal{T}$ can be pruned safely.

For example, in \Cref{fig:arborist} (b), the left tree containing point $p$ is discarded, whereas the middle tree containing point $q$ is retained since it may still yield a shorter path. The planner terminates once the goal $\mathcal{G}$ is reached and all other trees have been pruned. This procedure guarantees global optimality while minimizing redundant computation, as only trees with a provable potential for improvement are maintained. Our experimental results in \Cref{subsubsec:exp-cross-tree-check} validate this design. 

\subsection{Putting it Together}

\begin{algorithm}
\caption{\Name Motion Planning Algorithm}\label{alg:arborist}
\begin{algorithmic}[1]
\Require samples in $\mathcal{X}_{free}$, start $x_{init}$, goal $x_{goal}$, max trees $N$
\State Initialize: $V_{unvisited}\!\gets$ all samples; create tree $\mathcal{T}_0$ rooted at $x_{init}$ in manager $\mathcal{M}$; $reached\!\gets$ false
\While{true}
    \State $\mathcal{Z}\gets$ lowest-cost wavefront nodes across all active trees
    \State \textbf{if} $\mathcal{Z}=\emptyset$ \textbf{then return} success if $reached$ else fail
    \For{each $z\in\mathcal{Z}$ (with home tree $\mathcal{T}$)}
        \State \textbf{if} $z=x_{goal}$ \textbf{then} $reached\gets$ true
        \State \textbf{if} $reached$ \textbf{and} \Call{CrossTreeOptCheck}{} \textbf{then return} success
        \State \emph{Expand $z$:} for each $x\in$ \Call{PropagatingPoints}{$z$}, select the minimum-cost candidate parent from eligible OPEN neighbors; if its edge to $x$ is collision-free, apply \Call{ResolveClaim}{$x$} and move newly connected $x$ to $\mathcal{T}.V_{open}$
        \State Move $z$ from $V_{open}$ to $V_{closed}$
        \State \textbf{if} $|\mathcal{M}|<N$ \textbf{then} attempt split on $\mathcal{T}$
    \EndFor
\EndWhile
\end{algorithmic}
\end{algorithm}

\Cref{alg:arborist} summarizes the complete workflow of the proposed \Name algorithm in \Cref{fig:arborist}.
\Name breaks the inherently sequential execution of baseline FMT* by enabling concurrent multi-tree expansion. Through its coordinated mechanisms (tree overriding, global competition management, and cross-tree optimality checking), \Name preserves the dynamic-programming semantics of baseline FMT* and guarantees the same optimal path cost while significantly improving computational efficiency.

\noindent\textbf{Correctness intuition.} \Name preserves FMT* correctness because all introduced parallelism is \emph{schedule-only}. Lower-cost claims always override higher-cost ones, hibernation defers but never discards potentially optimal work, and termination removes only trees whose admissible lower bound cannot improve the incumbent solution. Taken together, these properties make \Name equivalent to a reordered sequential execution of FMT*. \Cref{fig:path_quality,fig:termination_ablation} provide empirical support for this intuition.

\noindent\textbf{Correctness invariant.} Consider the fixed sampled roadmap $G=(V,E)$ with deterministic edge validity and weights. Each node $v$ maintains a label $g(v)$, parent $p(v)$, and owner tree. A tree expansion only generates relaxation claims of the form $(x, g(y)+w(y,x), y, tree\_id)$ for collision-free edges $(y,x)$. These claims are committed through a hardware Commit Arbiter: for each node $x$, the arbiter selects the minimum-cost claim, with a deterministic tie-breaker, and updates the Node Table only if the claim improves $g(x)$. Since this operation is an atomic \emph{min} over valid relaxations, simultaneous claims to the same node are serializable and equivalent to a deterministic sequential order. Tree splitting only partitions the OPEN frontier. Tree override occurs only after a lower-cost claim wins and preserves the victim frontier under the winning owner. Therefore, by induction over committed relaxations, the parallel algorithm preserves the sequential relaxation semantics and converges to the same node labels; equal-cost ties can change only the selected parent, not the path cost.

To evaluate \Name on a conventional platform, we implement a multi-threaded C++ version and run it on a multiprocessor CPU. As shown in \Cref{tab:latency_matrix_fit}, four CPU threads provide $2.3\times$ speedup on 2D-City and $1.3\times$ on 2D-Maze relative to one thread. 
This limited gain suggests that general-purpose processors cannot effectively capture inter-tree parallelism, largely due to thread synchronization overhead, irregular execution, and memory bottlenecks in neighbor search. 
%These limited gains are consistent with thread synchronization overhead, irregular execution, and memory bottlenecks in neighbor search. 
These observations motivate a domain-specialized architecture for \Name. Accordingly, \Cref{sec:hardware} presents \Accel, a hardware design that enables \Name to better exploit both its inter-tree parallelism and additional architectural optimization opportunities for higher power, performance, and area (PPA) efficiency.

\section{\Name~Hardware Architecture}
\label{sec:hardware}

%\yt{Arborist -> inter-tree parallelism, speculation -> intra-tree parallelism, OUTLINE}

%\jeff{@Yaotian, can you use the following three papers as example to set up this section?}
% Frist paper, See Sec 5 and Fig.9
%https://dl.acm.org/doi/10.1145/3695053.3731110

% Second paper, see Section 4.2, Fig.6
%https://dl.acm.org/doi/10.1145/3669940.3707258

% Third paper, see Section 6, Fig.10
%https://dl.acm.org/doi/pdf/10.1145/3676641.3716016

\begin{figure}[t]
\centerline{\includegraphics[width=\columnwidth]{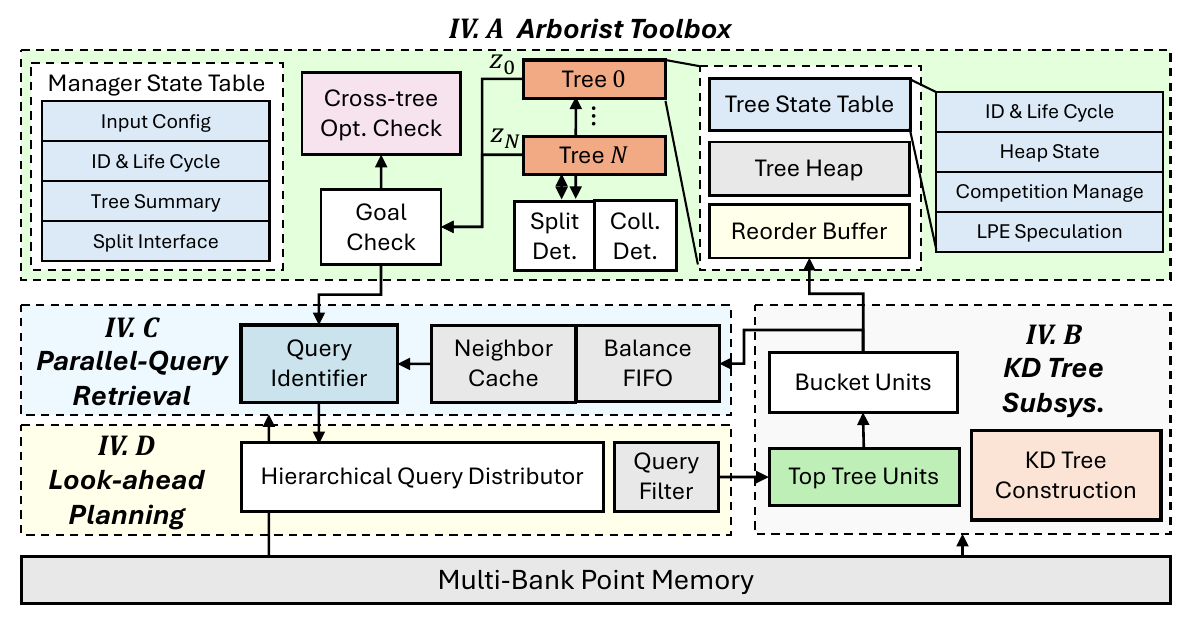}}
    \caption{Overview of \Accel. The architecture is organized into four components: the \Name Toolbox for multi-tree control, the KD-tree subsystem for neighbor search and construction, the Parallel-Query Retrieval path for balanced memory access and query generation, and the Look-ahead Planning path for speculative execution. These components interact through the multi-bank point memory to exploit both inter-tree and intra-tree parallelism.}
    \label{fig:arch}
\end{figure}

In this section, we present \Accel, our hardware architecture that implements \Name, as illustrated in \Cref{fig:arch}. \Accel is organized into four key components:
(1) the \textbf{\Name Toolbox} (\Cref{subsec:arborist_toolbox}), which provides hardware support for inter-tree parallelism described in \Cref{sec:arborist};
(2) a \textbf{KD-tree Subsystem} (\Cref{subsec:kd_tree_accel}) for high-throughput near-neighbor search and dynamic replanning support;
(3) a \textbf{Parallel-Query Retrieval (PQR)} module (\Cref{subsec:paral_query_retrieval}) that balances multi-bank memory accesses to alleviate system-level memory bandwidth bottlenecks; and
(4) a \textbf{Look-ahead Planning Engine (LPE)} (\Cref{subsec:deep_speculation}) that exploits intra-tree parallelism while preserving path quality by in-order commit. We further discuss the programmability and extensibility in \Cref{subsec:programmability}.

%We put auxiliary components as Other Design Considerations in \Cref{subsec:others}.
%\jeff{For jeff: carefully re-think each hardware blocks (and description) below for missing details or clarity}

\subsection{\Name Toolbox} 
\label{subsec:arborist_toolbox}

%\yt{draw a figure of split flow}

As shown in \Cref{fig:arch}, the \Name Toolbox is the control substrate that realizes the inter-tree parallelism introduced in \Cref{sec:arborist}. Rather than accelerating nearest-neighbor search directly, it manages the lifecycle of multiple concurrent FMT-style trees and enforces the control rules needed to preserve correctness under parallel expansion. Concretely, the toolbox consists of a global \Name Manager, a Split Detector, Tree Units, and a collision detector.
% Together, these modules implement tree creation, split coordination, ownership updates, hibernation/reactivation, and cross-tree termination checking.

\noindent\textbf{\Name Manager.}
%The \Name Manager keeps a table in \emph{registers} which tracks per-tree state, records the current path cost for each tree to support fast \textit{ResolveClaim} function call and \textit{Cross-tree Optimality Check}. It alsoimplements FSM-based control logic to orchestrate \texttt{tree-splitting} and \texttt{tree-overriding} operations. 
% The \Name Manager maintains a per-tree state table in registers, tracking current path costs and supporting fast execution of the \textit{ResolveClaim} and \textit{Cross-tree Optimality Check} operations. It also employs FSM-based control logic to orchestrate tree-splitting and tree-overriding events, coordinating inter-tree interactions across the system.
The \Name Manager is a lightweight global controller that maintains a manager state table and compact per-tree summaries, including tree identity, lifecycle state, and status signals needed for competition management and termination. It is responsible for selecting active trees, issuing split commands, coordinating node migration between trees, and arbitrating events such as override, hibernation, reactivation, and goal detection.

\noindent\textbf{Split Detector.}
The Split Detector is the hardware realization of the Step-1 gap test. Instead of sorting projected coordinates, it constructs a fixed-bin histogram for each examined dimension, converts non-empty bins into an occupancy bitset, and identifies the Longest Zero Run (LZR). A zero run of length LZR corresponds to an empty coordinate interval of approximately $\mathrm{LZR} \cdot w_{\mathrm{bin}}$. The detector uses the threshold
\( T_h \coloneqq \beta\, r_n / w_{\text{bin}} \)
and triggers when \(\mathrm{LZR} \ge T_h\), where \(w_{\text{bin}}\) is the bin width, \(r_n\) is FMT*’s neighbor radius, and $\beta$ is a tunable hyperparameter. This approximates the software condition $G_d \ge \beta r_n$. The lightweight detector enables real-time identification of separable clusters with minimal computational overhead.

% \begin{figure}[tbp]
% % \vspace{-10pt}    
% \centerline{\includegraphics[width=0.8\columnwidth]{figs/split.pdf}}
%     \caption{\Name Toolbox Module. }
%     \label{fig:toolbox}
% \end{figure}

\noindent\textbf{Tree Units.}
Each active tree unit is backed by a dedicated heap that stores its open-set wavefront as an independent priority queue. These heaps are paired with per-tree state tables that record local execution metadata, such as heap status and tree-level bookkeeping needed by the manager. Each tree unit also has a reorder buffer to support LPE speculation mentioned in \Cref{subsec:deep_speculation}.

\noindent\textbf{Collision Detector.} \Accel exposes collision checking through a fixed edge-validity request/response interface rather than a universal rule language. The FMT* datapath issues a request containing the candidate edge endpoints and receives a binary valid/collision response. For 2D/3D workloads, the checker performs occupancy-grid or voxel-grid traversal over map memory. For 5D/7D arm workloads, it performs fixed-DoF forward kinematics and rasterizes each link against a 2D occupancy map. Because FMT* performs lazy edge validation and collision checking is a small fraction of the profiled runtime, \Accel focuses on the shared FMT substrate; retargeting binds the same interface to the appropriate benchmark-specific checker.
% \noindent\textbf{Collision Detector.}
% Because FMT* performs lazy edge validation, collision checking is a minor cost in our workload profile (\Cref{fig:comb_profiling}). \Name therefore employs lightweight Collision Detection Units that perform the benchmark-specific edge-validity test: obstacle-map checks for 2D/3D workloads, and benchmark-defined connection-validity checks for 5D/7D arm workloads. Each unit returns a binary valid/collision result.
 
%$Th.$ is the split threshold. If the LZR is higher than the threshold, the Manager will starts to pop the splitted nodes from heap in-use to the vacant heap.
%Because multiple trees grow concurrently, we instantiate multiple heap units to store $V_{\mathrm{open}}$. 

\subsection{KD-Tree Subsystem with Dynamic Planning Support}
\label{subsec:kd_tree_accel}
As shown in \Cref{fig:arch}, \Accel combines a two-stage KD-tree Subsystem~\cite{EASTMAN1981165, zhou2008real, deBerg2008, pinkham_quicknn_2020, xu2019tigris} for near-neighbor queries with an exact-balanced KD-tree constructor for dynamic-environment support.

%\jeff{@yaotian: question: do we only need to get the nearest neighbor from KD-Tree, or we also need to get 2nd, 3rd near neighbors? Let's say for deep speculation case.} \yt{No, either radius search or K-NN search, both have multiple near points.}

\textbf{KD-Tree Search.}
\Accel{}'s KD-tree search engine builds on Tigris~\cite{xu2019tigris}, employing a two-stage pipeline: (1) multiple front-end Top-Tree Units (TTUs) that exploit query-level parallelism, and (2) multiple back-end Bucket Units (BUs) that provide node-level parallelism.

% However, the original Tigris design exhibits low hardware utilization and high cache miss rate under \Name’s irregular and sparse query patterns. To address this, \Accel co-designs the top-tree units and the bucket cache with a novel Parallel Query Retrieval engine to dynamically balances workload distribution across KD-tree query processing units, maximizing KD-tree subsystem utilization, 
% improving the bucket cache performance, and 
% sustaining high throughput even under bursty workloads, as detailed in \Cref{subsec:paral_query_retrieval}.

\textbf{Exactly Balanced Two-Stage KD-Tree Constructor for Fast Tree Build and Dynamic Planning.}
In software implementations, KD-tree construction accounts for only about 0.6--6.3\% of the total time, since tens of thousands of KD-tree searches dominate. 
%In the retained four-planner CPU summaries, KD-tree construction accounts for 0.6--6.3\% of total time. 
Under the \Accel optimization it exceeds 50\%, creating a need for an on-chip KD-tree constructor. 
%However, with the optimizations proposed here, construction time becomes significant (>50\%), creating a need for an on-chip kd-tree constructor.
Unlike Tigris~\cite{xu2019tigris}, \Accel integrates an on-chip KD-tree constructor that supports rapid re-sampling and balanced top-tree reconstruction. The constructor uses a hardware-optimized Quickselect algorithm~\cite{bentley1999programming} to identify the exact median in expected $O(n)$ time without sorting all samples. This design maintains balanced KD-trees across planning iterations, thereby preserving high neighbor-search throughput and predictable latency under dynamic workloads. It also enables the deployment for IAFMT*/IABFMT* on \Accel as they need iterative planning (\Cref{subsec:programmability}).

\subsection{Parallel-Query Retrieval}
\label{subsec:paral_query_retrieval}

The Parallel Query Retrieval (PQR) subsystem in \Accel optimizes query throughput, memory bandwidth, and access locality through three tightly integrated components: (1) a Ring-buffer Neighbor Cache for high-reuse query data, (2) a high-throughput query identifier coupled with a multi-banked point memory for concurrent access, and (3) a Balance FIFO for data layout optimization. Together, these modules sustain high utilization of the KD-tree subsystem and alleviate the memory bottlenecks that typically limit nearest-neighbor search performance.

\begin{figure}[t]
  \centerline{\includegraphics[width=1\columnwidth]{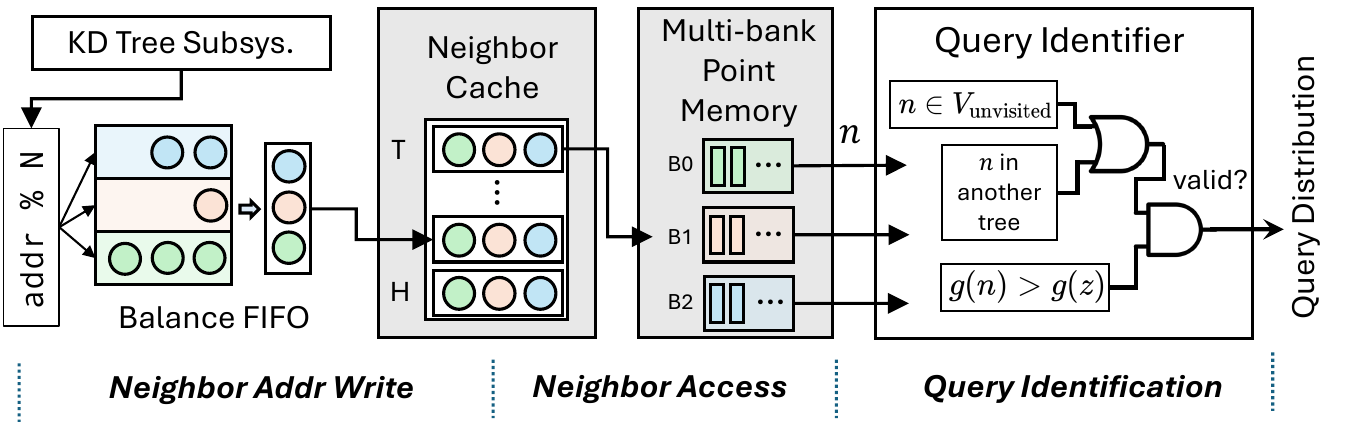}}
    \caption{Parallel-Query Retrieval design. Neighbor addresses from the KD-tree subsystem are balanced across FIFO lanes, cached, used to access the multi-bank Point Memory, and filtered by the Query Identifier to generate valid downstream queries. Circles denote addresses, squares denote node entries, and H and T denote the head and tail of the ring buffer.}
    \label{fig:paral_qid}
\end{figure}

As illustrated in \Cref{fig:paral_qid}, \Accel employs a \emph{Neighbor Cache} to store each returned neighbor's address together with the edge cost computed during KD-tree radius search. Caching the cost allows downstream stages to reuse the neighbor list without refetching point coordinates for distance recomputation. The cache is implemented as a \emph{ring buffer} to efficiently accommodate variable-length neighbor lists produced during each motion planning iteration.
A Query Identifier then fetches the corresponding neighbor points from the on-chip \emph{Multi-bank Point Memory} in response to neighbor cache reads. Points that pass the \textsc{PropagatingPoints} check (\Cref{subsec:competition-management}) are marked as valid queries and dispatched to the next stage.

\noindent\textbf{Multi-banked Point Memory.}
% Empirically, fewer than 10\% of queries are valid. 
To sustain sufficient throughput, \Accel employs a multi-banked Point Memory design that provides high read bandwidth and ensures an adequate number of valid queries per cycle. This architecture significantly improves KD-tree TTU utilization.

\noindent\textbf{Balance FIFO for Data Layout Optimization.}
\label{subsubsec:redirect_FIFO}
%better data layout to reduce bank read conflict.
However, simply increasing the number of point-memory banks does not guarantee full throughput because neighbor counts vary across query nodes, producing a nonuniform data layout across banks.
To address this, we insert a lightweight Balance FIFO (BF), inspired by~\cite{ausavarungnirun2012staged}, into the neighbor-cache write stage. It dynamically equalizes per-bank load before neighbor addresses are enqueued in the Neighbor Cache. Each BF lane is aligned with a Point Memory bank. In practice, BF delivers substantial speedups by redistributing neighbor addresses and balancing traffic to the Point Memory subsystem (see \Cref{subsubsec:rf_vs_os}).

% one conventional approach is \emph{over-sampling (OS)} \cite{jegou2011searching, mslearn_quantization_vector_search}, which proactively reads multiple address batches (from the neighbor cache), re-mapping them for a better-balanced data layout. While sound, OS pressures the neighbor cache memory reads and requires a crossbar.
% %introduces read-side latency and complexity. 

% In \Accel, we address this issue earlier, i.e., during the neighbor cache write stage, by introducing a lightweight Redirect FIFO (RF) plug-in that monitors each FIFO lane status and dynamically equalizes per-bank load before neighbor addresses are enqueued in the cache (\Cref{fig:our_near_cache}). 
% Each RF lane is aligned with a Point Memory bank.
% In practice, RF delivers substantial speedups by suppressing randomness in neighbor node distribution and driving balanced traffic to the Point Memory subsystem (see \Cref{subsubsec:rf_vs_os}).

\subsection{Look-ahead Planning}
\label{subsec:deep_speculation}

\Accel supports intra-tree parallelism through a speculative execution strategy that breaks the inherent data dependencies across iterations in FMT*, realized by the Look-ahead Planning Engine (LPE).

\noindent\textbf{Intuition.}
FMT* expansion is inherently sequential: each wavefront of nodes must be fully processed before the next can begin. This strict ordering limits hardware utilization, as many compute units remain idle while waiting for dependencies to resolve. However, most node expansions in practice are independent; their outcomes do not affect each other’s feasibility or cost evaluation within the same iteration.

\begin{figure}[t]
 \centerline{\includegraphics[width=\columnwidth]{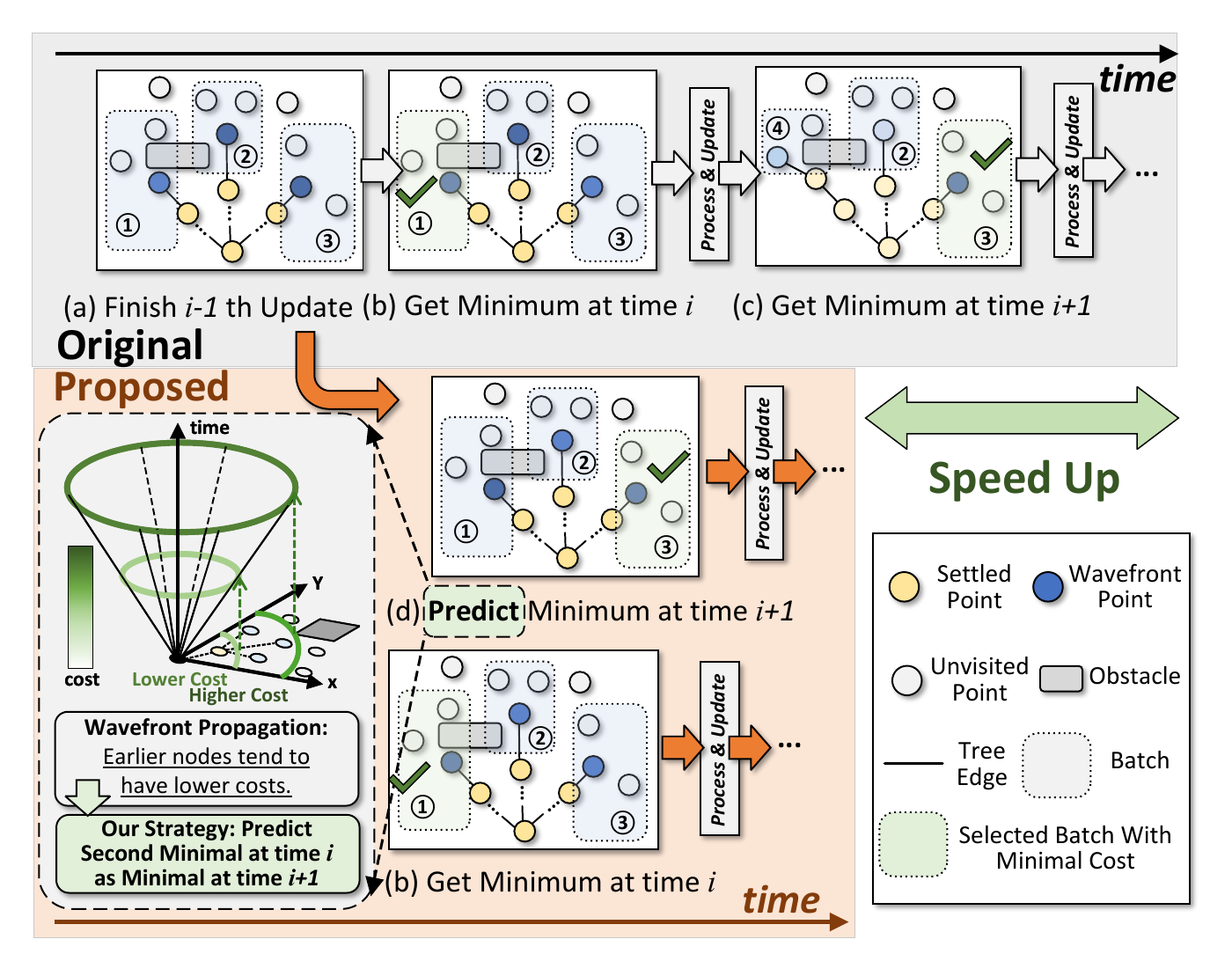}}
    \caption{Example of Look-ahead Planning ($L=1$).
    In the baseline pipeline (top), steps (a), (b), and (c) execute sequentially: in (b), the algorithm selects the minimum-cost wavefront point at time~\(i\), and only after processing and updating it can the minimum for time~\(i{+}1\) be determined.
    The proposed scheme (bottom) predicts that earlier-generated wavefront nodes tend to have lower costs, as shown in the bottom-left inset~\cite{Sethian2006FastMarching}. It can therefore process the minimum at time~\(i\) in (b) and the predicted minimum at time~\(i{+}1\) in (d) in parallel.
    }
    \label{fig:speculation}
    %\vspace{-10pt}
\end{figure}

\begin{figure}[t]
 \centerline{\includegraphics[width=\columnwidth]{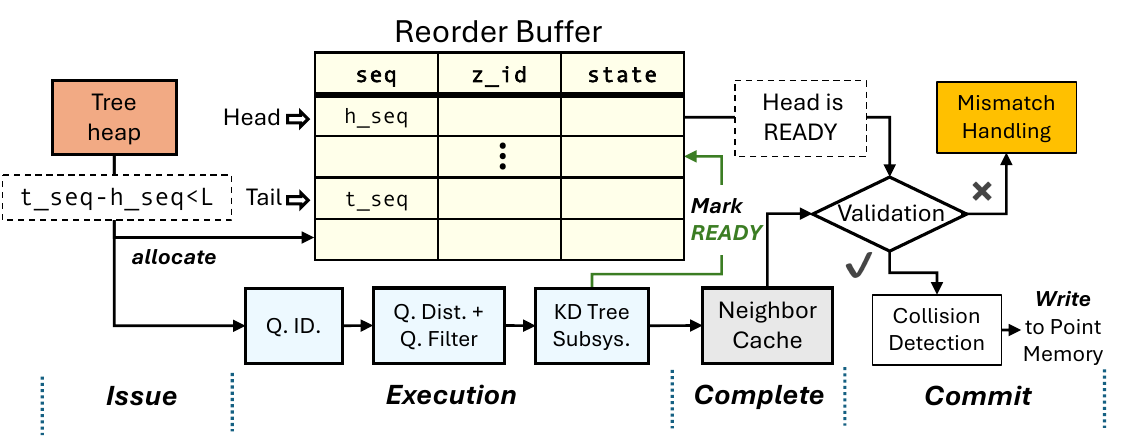}}
    \caption{LPE architecture. A per-tree reorder buffer (ROB) enables speculative look-ahead execution of future wavefront nodes while preserving sequential FMT* semantics through in-order commit.}
    \label{fig:lp_arch}
\end{figure}

\noindent\textbf{Design.}
As shown in \Cref{fig:speculation}, the Look-ahead Planning Engine (LPE) speculatively advances future wavefront expansions, i.e., up to $L$ look-ahead steps, before the current iteration is fully committed. To preserve path quality and correctness under speculation, each tree unit is equipped with a lightweight Reorder Buffer (ROB) that enforces \emph{in-order commit} of valid wavefront queries, thereby preserving the semantics of sequential FMT* execution. The physical ROB storage is implemented as a circular buffer indexed by the sequence number modulo the ROB capacity, with head and tail counters used for commit and allocation.

The LPE microarchitecture is illustrated in \Cref{fig:lp_arch}. Similar to a conventional out-of-order processor, it is organized into four stages: \emph{issue}, \emph{execution}, \emph{complete}, and \emph{commit}, all centered around the ROB. During the issue stage, a new wavefront node $z$ can be issued from the tree heap and allocated into the ROB whenever the speculation window is not full, i.e., when \texttt{t\_seq - h\_seq < L}. Here, \texttt{h\_seq} and \texttt{t\_seq} denote the sequence numbers of the ROB head and tail, respectively. 

In the execution stage, the issued node passes through the Query Identifier, Query Distributor, and Query Filter and then enters the KD-tree subsystem to retrieve its neighbor information. In the complete stage, the retrieved neighbor list is written into the Neighbor Cache, and the corresponding ROB entry is marked \texttt{READY}. Once the head entry becomes \texttt{READY}, it proceeds to the commit stage.

At commit time, the ROB validates whether the speculative result is still consistent with the \emph{current} tree state. Specifically, when committing $z_j$, for each candidate $x$ produced during speculation, we recompute its best parent using the current open set:
\begin{equation}
\label{eq:speculative_check}
y^*(x) \in \arg\min_{y\in V^{(j-1)}_\mathrm{open} \cap \mathrm{Near}(x)} \phi_y(x),
\end{equation}
where $\phi_y(x)\coloneqq g(y)+c(y,x)$ is the candidate cost-to-come through $y$, and $c(y,x)$ is the connecting edge cost.
ROB validation reuses the cached neighbor entries generated during speculative execution. For each cached neighbor $y$, the validation stage reads the corresponding node-state entry from the per-tree Tree State Table to obtain current OPEN membership and cost-to-come $g(y)$, filters non-OPEN nodes, and reduces $\phi_y(x) \coloneqq g(y)+c(y,x)$ using the cached edge weight. This avoids scanning the OPEN heap, scanning the Tree State Table, rerunning KD-tree search, or refetching full point coordinates during commit validation.
We then perform collision checking and update the tree state only at commit. In this way, speculation is allowed to compute read-only intermediate results early, but it is not allowed to modify the global planning state before validation.

If earlier commits have changed the wavefront such that the speculative choice is no longer optimal or feasible, i.e., \Cref{eq:speculative_check} fails, the \emph{Mismatch Handling} module is triggered. It flushes all younger ROB entries according to their \texttt{seq} numbers, reinserts the flushed $z$ nodes back into the tree heap, and restarts execution from the corrected state.

% This design allows LPE to overlap multiple dependent FMT* iterations while using the ROB to recover the original sequential ordering at commit.

\begin{figure}[t]
\centerline{\includegraphics[width=\columnwidth]{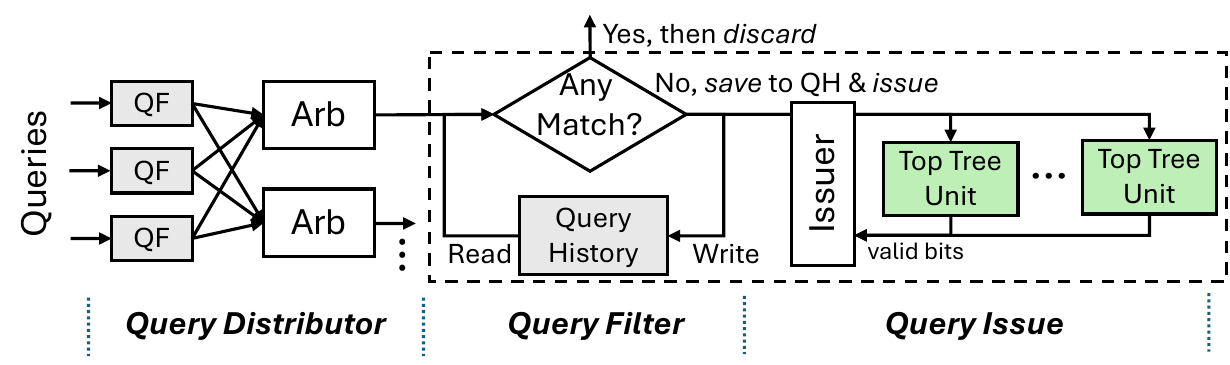}}
\caption{Hierarchical Query Distributor with Query Filter. Queries are buffered in Query FIFOs (QFs), selected by lightweight arbiters, checked against a per-set query-history structure, and then issued to available Top Tree Units. Duplicate queries are discarded before issue, reducing redundant speculative KD-tree lookups.}
\label{fig:query_distributor}
\end{figure}

\noindent\textbf{Hierarchical Query Distributor with Query Filter.}
\label{subsubsec:query_distributor}
With \Accel's Look-Ahead Planning and Tree-Splitting mechanisms, the KD-tree subsystem must sustain a substantially higher query rate. Supporting both inter-tree and intra-tree parallelism requires issuing multiple queries per cycle, making query dispatch a key bottleneck. To address this challenge, \Accel employs a \emph{Hierarchical Query Distributor with Query Filter} (HQD-QF), shown in \Cref{fig:query_distributor}, which partitions the Top Tree Units (TTUs) into independent sets and performs scalable query dispatch with early duplicate suppression.

The HQD-QF is organized into three stages. First, incoming queries are buffered in multiple Query FIFOs (QFs). Second, a set of lightweight arbiters selects valid queries from these QFs and forwards them to their assigned TTU set. Third, before issue, each query is checked against a lightweight per-set \emph{Query History} structure that records recently issued queries. If a match is found, the query is discarded; otherwise, it is inserted into the query history and forwarded to a local issuer, which dispatches it to an available TTU according to the returned valid bits.

% This hierarchical organization replaces a single global dispatcher with multiple local arbitration domains, thereby eliminating costly global arbitration, shortening the critical path, and scaling efficiently to large TTU arrays under heavy query traffic. At the same time, the Query Filter removes redundant speculative queries early, reducing unnecessary KD-tree activity and improving effective TTU utilization with minimal hardware overhead.

% \subsection{Other Modules}
% \label{subsec:others}

% The \emph{Find Min Connect} ($Min(\cdot)$) and \emph{Collision Detection} (CD) modules are duplicated to support \Name's $Max_{num}$ trees. $Min(\cdot)$ includes a comparator and registers to store the connection point with the lowest path cost. 
% Note that CD is not a major bottleneck of FMT*.
% \Accel uses a group of linear-feedback shift registers for sampling.

% %\rule{\columnwidth}{0.1pt}
% %\jeff{@yaotian, help add one or two sentences describe this figure here}

% Together, the \Accel forms a tightly integrated pipeline that maximizes inter- and intra-tree parallelism while preserving correctness. The following section details \Accel’s implementation and evaluates its architectural efficiency across diverse motion-planning workloads.

\subsection{Programmability and Extensibility}
\label{subsec:programmability}

\begin{figure}[t]
\centerline{\includegraphics[width=\columnwidth]{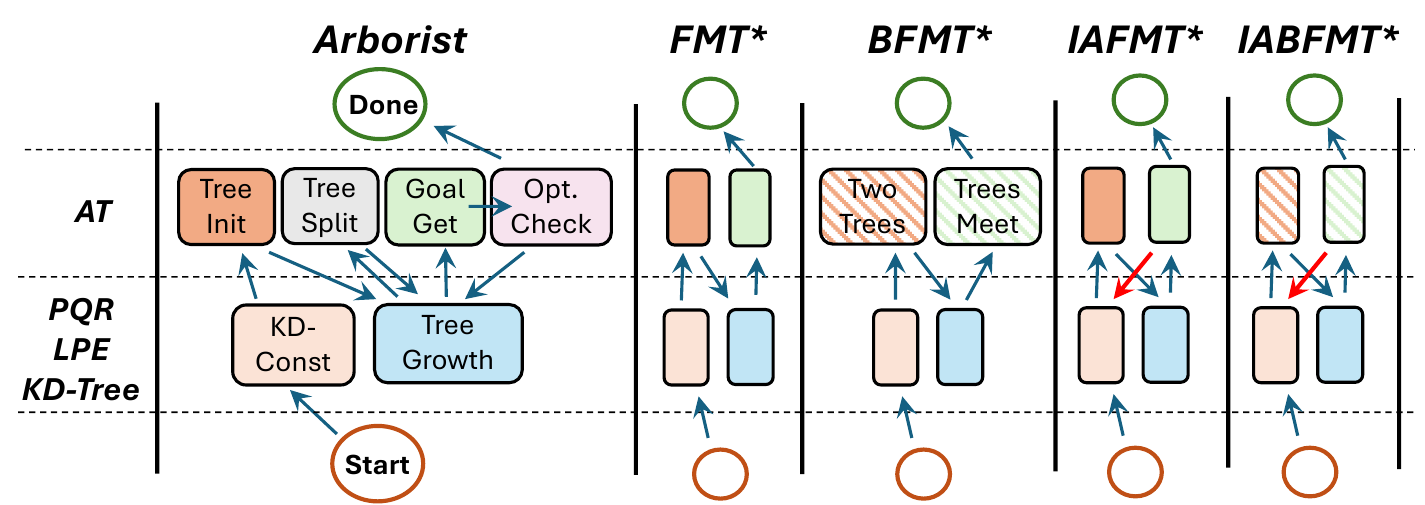}}
\caption{Configurable \Accel organization for FMT-family planners. 
All variants reuse the PQR/LPE/KD-tree execution kernel, while planner-specific behavior is realized by reconfiguring the control layer. Modules with same color are same config directly reuse.
%The fixed/model evaluation reuses the PQR/LPE/KD-tree execution model while replaying planner-specific software control flows;
Striped modules denote intended control changes and the red path denotes iterative planning feedback.}
\label{fig:extensibility}
\end{figure}

\noindent\textbf{FMT-Family Support.}
Although \Accel is built from specialized functional units, it is not hard-wired to a single planner. Its flexibility comes from a clean separation between a \emph{fixed execution substrate} (the KD-tree pipeline, PQR, and LPE, which together implement the kernels shared by FMT-family planners: neighbor retrieval, candidate-parent evaluation, lazy edge validation, and ordered commit) and a \emph{configurable control layer} exposed via the \Name Toolbox (active-tree count, root initialization, split threshold, goal/meeting logic, override/hibernation, pruning, and look-ahead depth). Future algorithmic changes that preserve the FMT-style wavefront expansion can therefore be realized by reconfiguring control policies while reusing the same datapath.

\Cref{fig:extensibility} illustrates this decomposition: the upper row represents planner-specific control behavior, while the lower row represents the shared PQR/LPE/KD-tree module. FMT*, BFMT*, IAFMT*, and IABFMT* exercise different software control paths; \Name enables tree splitting, overriding/hibernation, and cross-tree optimality checking. As a result, the same silicon retargets across
the FMT*-family without redesigning the core neighbor-search, memory, or tree-expansion datapath.

\noindent\textbf{Arbitrary Dimension Support.} We support multidimensional states using a fixed-width vector datapath and hardware time-multiplexing. The datapath processes four dimensions (D) per cycle. If $D>4$, \Accel reuses the datapath across multiple cycles.

\section{Evaluation Setup}
This section details \Name's implementation, together with the workloads, baselines, and metrics used for our evaluation.
\subsection{\Name Implementation}

%\jeff{@Yaotian, polish it refer to following example}\yt{filled out}
%1. Section 5 in https://dl.acm.org/doi/10.1145/3669940.3707258

We implement both the original FMT* and \Name algorithm in C++. 
\Accel is written in SystemVerilog and synthesized with Synopsys Design Compiler on a commercial 16nm technology process, 
with a target frequency of 500 MHz. All memory modules are implemented with foundry-provided SRAM macros.

\noindent\textbf{Hardware Configuration.}
For the \Name Toolbox, we instantiate one \Name Manager, one Split Detector, and eight per-tree heaps (each with a \SI{2}{\kilo\byte} SRAM).
The KD-tree subsystem includes \SI{64}{} TTUs and \SI{64}{} BUs, backed by a \SI{2}{\kilo\byte} top-tree cache and a \SI{1}{\kilo\byte} bucket cache.
The LPE uses a \SI{128}{\byte} reorder buffer.
The PQR subsystem comprises a 32-bank Point Memory (\SI{256}{\kilo\byte} total) and a \SI{16}{\kilo\byte} Neighbor Cache; the Balance FIFO threshold is set to 8 entries, with a total size of \SI{768}{\byte}. The design has eight Min. Connect Units and eight Collision Detection Units.
%\jeff{@yaotian, what is the total power and area for this design?}

%\noindent\textbf{Accelerator Power and Area.} The retained component summary totals \SI{379.6}{\mW} and \SI{1.22}{\mm\squared} after paper rounding.

\noindent\textbf{Accelerator Power and Area.} The total power and area of \Accel are \SI{379.6}{\mW} and \SI{1.22}{\mm\squared}, respectively.

\noindent{\textbf{Latency.} The latency numbers are from our cycle-accurate ArboristAccel simulator. It explicitly accounts for dynamic effects including speculation success/failure, false-speculation recovery through the LPE mismatch-handling path, tree overriding/hibernation, variable neighbor-list lengths, and multi-bank memory conflicts.}

%\noindent\textbf{Simulation Methodology.}
%check Section 5 in https://dl.acm.org/doi/10.1145/3695053.3731003
%We simulate the entire system with our cycle-accurate, event-driven simulator, implemented with component-level latencies. We also record the active cycles for essential computing units and memories for a more accurate power estimation.

\subsection{Workloads}
%\lingyi{Consider to change the title to "Workloads"? also more explanation on why these workloads?}
%\jeff{@01, Yes, can you help insert one or two sentences with your color?}
% \begin{comment}
% \Name is evaluated on RTRBench~\cite{rtrbench}. 
% \yt{RTRBench is also used by RACOD and MOPED, previous path planning accelerators}
% For 2D environments, we select city and maze maps from \cite{sturtevant2012benchmarks}, while 3D environments are drawn from~\cite{octomap}. In addition, we use the 5D Robot Arm provided by RTRBench and extend it to a 7D Arm for higher-dimensional evaluation.
% \end{comment}

\Name is evaluated on RTRBench~\cite{rtrbench}, a standardized motion-planning benchmark suite that is also used by prior path-planning accelerators such as RACOD~\cite{bakhshalipour2022racod} and MOPED~\cite{huang2024moped}.
%\jeff{@yaotian, add refs} \yt{DONEf}
We use a set of five representative workloads that span different application domains and planning dimensionalities.
For 2D navigation, we select city and maze maps from the widely used benchmark suite of~\cite{sturtevant2012benchmarks}, covering urban driving–like scenarios as well as highly cluttered maze environments.
For 3D planning, we adopt volumetric maps from OctoMap~\cite{octomap}, representative of UAV and aerial-robot workloads.
To evaluate high-DoF manipulation, we follow RTRBench and use its 5D robot arm and an extended 7D arm configuration.

% \begin{comment}
% \subsection{Metrics}
% We evaluate \Name on path planning quality, latency, computation cost, power consumption, and area. 
% Here, path quality refers to the geometric optimality of the resulting trajectory and, in this work, is measured by the path length, which is a critical metric because it directly decides the time and energy required for the robot to complete its task. 
% Computation cost is defined as the number of queries processed. 
% Latency and power are derived from a cycle-accurate simulator, which models execution cycles and memory access patterns. 
% Area results are obtained from post-synthesis reports. 
% Area efficiency is defined as for each unit area, how much speedup is obtained. Power efficiency is defined as for each unit power (watt), how much speedup is obtained.

% %\jeff{@01 or Yaotian, can you help explain what are speedup, area efficiency, energy efficiency in Fig.14?} \yt{speedup is explained in 6.1, as CPU to be the baseline}

% %\jeff{@yaotian, check if there are any other metrics in Sec 6 we didn't explain here.}  \yt{added area and energy efficiency}
% \end{comment}

\subsection{Hardware Baselines}
We compare \Name against four baselines:
\begin{itemize}[leftmargin=1em]
 \item \textbf{CPU \Name}: a multi-thread C++ implementation of \Name compiled with \texttt{-O3}, measured on an AMD Ryzen Threadripper 3960X (3.8\,GHz, 768\,GB RAM). Area/power are reported on 4 cores (50\,mm$^2$, 40\,W).
 \item \textbf{GPU GMT*}~\cite{GMT}: 
 a GPU implementation of FMT* with a 10\% path-quality drop; included as a high-performance baseline. Performance is taken from the reported NVIDIA GTX 980 results (398\,mm$^2$, 165\,W TDP).
 %We estimate each workload's latency as $0.01\times$ its single-thread CPU latency, motivated by the approximately $100\times$ acceleration reported for GMT* on an NVIDIA GTX 980. These are scaled estimates, not per-workload GPU measurements. We use 398\,mm$^2$ die area and 165\,W TDP. The published GMT* experiments report approximately 10\% higher path cost; this quality difference is not independently measured on our workloads.
 \item \textbf{ASIC w/ Tigris}~\cite{xu2019tigris}: 
 Our baseline FMT* ASIC design based on Tigris-style KD-tree. 
 %executed by Tigris's KD-tree accelerator. 
 We synthesize it with 500 MHz on the same 16nm node for a fair comparison (0.635\,mm$^2$, 134.5\,mW).
 %An author-modeled FMT* ASIC using a Tigris-style KD-tree; this is not a published Tigris FMT result.
 %executed by Tigris's KD-tree accelerator. 
 %The retained model uses a 500\,MHz target and fixed 16\,nm component sums of 0.635\,mm$^2$ and 134.5\,mW.
 \item \textbf{MOPED}~\cite{huang2024moped}: a state-of-the-art motion planning accelerator based on RRT*~\cite{karaman2011sampling}. The results are scaled to 14 nm using Deep-ScaleTool~\cite{sarangi2021deepscaletool}, since 16 nm is not directly supported; 14 nm is a favorable scaling to MOPED and therefore conservative with respect to our claimed improvement. The original metrics at 28 nm are 0.62 mm$^2$ and 137.5 mW. The latency is obtained by running RRT* on same test cases until the path costs match our results with FMT*/Arborist, then scale to MOPED using the speedup factor ($\sim3000\times$) reported in~\cite{huang2024moped}.
\end{itemize}

% \begin{table}[h]
%   \centering
%   \caption{\hl{Area and power comparison} \textcolor{red}{To save space, delete this table, and just add Area and power numbers to each baseline.}}
%   \label{tab:area_power}
%   \begin{tabular}{lcc}
%   \toprule
%   Config & Area (mm$^2$) & Power (mW) \\
%   \midrule
%   CPU              & 296.0 & $4.0 \times 10^4$   \\
%   GPU GMT*         & 398.0 & $1.65 \times 10^5$ \\
%   MOPED            & 0.620   & 137.5    \\
%   Ours             & 1.043   & 304.0    \\
%   \bottomrule
%   \end{tabular}
%   \end{table}

Unless stated otherwise, \Cref{tab:scene-params} lists the \Name evaluation settings. $N$ is the number of samples; $\beta$ controls the LZR Split Detector; \texttt{MAX\_TREE} is the maximum number of trees; and $L$ is the number of look-ahead steps.

\begin{table}[thbp]
  \centering
  \caption{Evaluation Settings.}
  \label{tab:scene-params}
  \footnotesize
  \setlength{\tabcolsep}{4pt}
  \renewcommand{\arraystretch}{0.9}
  \begin{tabular}{lccccc}
  \toprule
  Scene & Dim & $N$ & $\beta$ & \texttt{MAX\_TREE} & LPE Step $L$ \\
  \midrule
  2D-City  & 2 & 20k & 0.75 & 8 & 5 \\
  2D-Maze  & 2 & 20k & 0.50 & 8 & 5 \\
  3D-Drone & 3 & 20k & 0.50 & 8 & 10 \\
  5D-Arm   & 5 & 10k & 0.50 & 8 & 5 \\
  7D-Arm   & 7 & 10k & 0.25 & 8 & 10 \\
  \bottomrule
  \end{tabular}
\end{table}

\section{Evaluation}

We first evaluate \Name's system-level benefits and then examine the impact of each architectural decision on \Accel performance.

\noindent\textbf{Efficiency Definition.} We define power and area efficiency as $E = \mathrm{speedup} \times (V_{baseline} / V)$, where $V$ represents the reported or modeled power or area, and speedup is relative to the four-thread CPU baseline. This metric penalizes designs that achieve speedup through disproportionate increases in power or area.

\subsection{System-level Benefits}

\begin{figure}[t]
  \centerline{\includegraphics[width=\columnwidth]{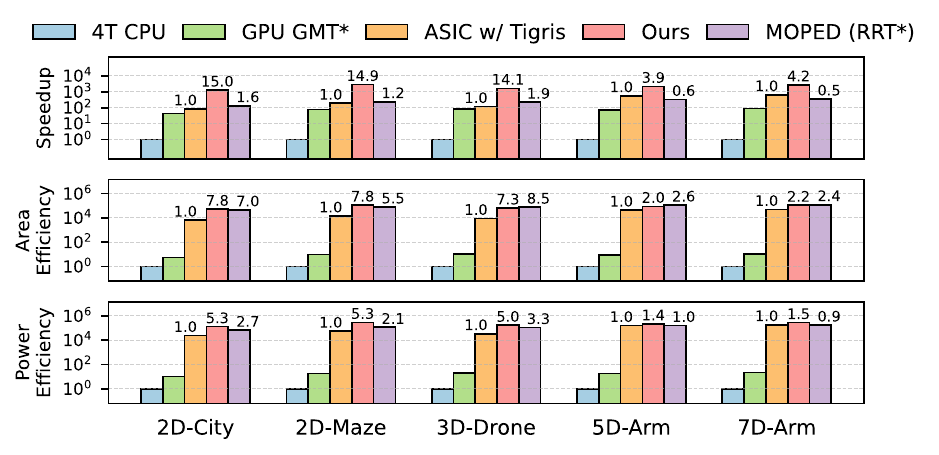}}
    \caption{System PPA comparison. Annotated values are normalized to ASIC w/ Tigris; the y-axis is normalized to the four-thread CPU baseline.}
    \label{fig:system_ppa}
\end{figure}

\noindent\textbf{Power, Performance, and Area (PPA) Comparison.}
Figure~\ref{fig:system_ppa} compares \Accel against CPU, GPU, ASIC w/ Tigris, and MOPED across five workloads using the performance model. In \textbf{speedup}, GPUs are around $10^{2}$, while ASIC w/ Tigris and MOPED peak at $6.4\times 10^{2}$ and $3.5\times 10^{2}$, respectively. \Accel is the \textbf{best performer overall}, reaching \textbf{$2.8\times 10^{3}$} on 2D-Maze and sustaining more than $10^{3}$ speedup across all workloads. Its \textbf{area} and \textbf{power efficiency} (normalized to 4T CPU) reach \textbf{$1.1\times 10^{5}$} and \textbf{$2.9\times 10^{5}$}, respectively. Averaged across workloads (geomean), \Accel improves over ASIC w/ Tigris by \textbf{$8.8\times$} in speedup, \textbf{$4.6\times$} in area efficiency, and \textbf{$3.1\times$} in power efficiency. Relative to MOPED, the corresponding factors are \textbf{$8.6\times$}, \textbf{$1.0\times$}, and \textbf{$1.8\times$}, respectively.
%Using the retained fixed component-power sum ($379.6$ mW), the modeled end-to-end energy per planning query remains below $1.2$ mJ across all workloads.
The end-to-end energy per planning query remains below $1.2$ mJ across all workloads on \Accel. 
Specifically, \Accel consumes $0.590$ mJ on 2D-City, $1.175$ mJ on 2D-Maze, $0.763$ mJ on 3D-Drone, $0.400$ mJ on 5D-Arm, and $0.630$ mJ on 7D-Arm.

%\noindent\textbf{Algorithm Speedup with Multi-thread CPU.}

The execution latencies are shown in \Cref{tab:latency_matrix_fit}. We profile CPU \Name using up to 24 threads. Due to multithreaded synchronization overhead, CPU speedup largely saturates at four threads when processing parallel trees.

   \begin{table}[t]
    \centering
    \caption{Latency comparison (ms); T denotes CPU threads.}
    \label{tab:latency_matrix_fit}
    \small
    \resizebox{\columnwidth}{!}{%
    \begin{tabular}{lcccc!{\color{lightgray}\vrule}cccc}
    \toprule
     & \multicolumn{4}{c!{\color{lightgray}\vrule}}{CPU} & GPU & ASIC w/ &  &  \\
    Scene & 1T & 2T & 4T & 24T & GMT* & Tigris & MOPED & \textbf{Ours} \\
    \midrule
    2D-City  & 4,466  & 2,550 & 1,951 & 2,009 & 44.66 & 23.322 & 15.00 & \textbf{1.555} \\
    2D-Maze  & 11,298 & 8,044 & 8,668 & 7,831 & 112.98 & 46.276 & 37.95 & \textbf{3.096} \\
    3D-Drone & 4,025  & 3,695 & 3,275 & 3,026 & 40.25 & 28.261 & 14.88 & \textbf{2.009} \\
    5D-Arm   & 3,083  & 2,765 & 2,294 & 2,133 & 30.83 & 4.120 & 7.14 & \textbf{1.054} \\
    7D-Arm   & 5,010  & 5,461 & 4,441 & 4,509 & 50.10 & 6.951 & 12.80 & \textbf{1.660} \\
    \bottomrule
    \end{tabular}%
    }
  \end{table}

\begin{figure}[t]
  \centerline{\includegraphics[width=\columnwidth]{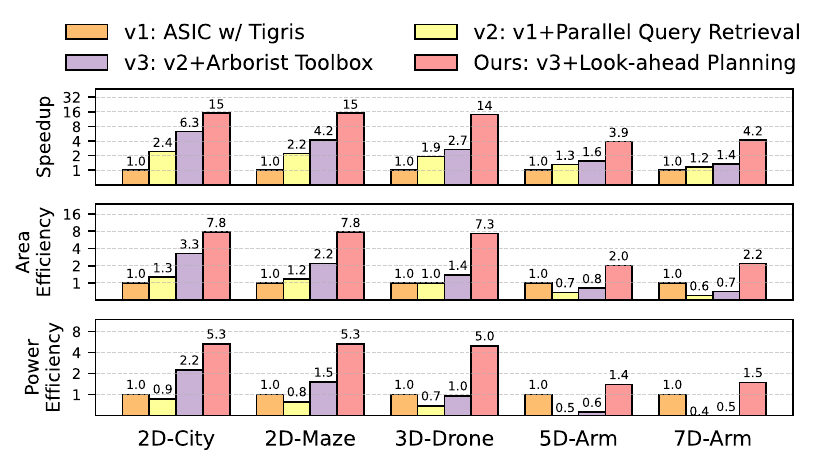}}
    \caption{Ablation study on \Accel.}
    \label{fig:ablation}
\end{figure}

\noindent\textbf{Ablation Study.}
\Cref{fig:ablation} reports an ablation study that quantifies the incremental contributions of \Accel's key innovations, i.e., Parallel Query Retrieval, \Name Toolbox, and Look-ahead Planning, to overall system efficiency.
%All the ``small'' optimizations discussed in the previous evaluation have already been applied in all cases.
Starting from the \textbf{ASIC w/ Tigris} baseline, adding \textbf{Parallel Query Retrieval} yields $1.2$--$2.4\times$ speedup and area efficiency up to $1.3\times$, but area efficiency falls below $1\times$ in 5D/7D. Incorporating the \textbf{\Name Toolbox} reaches $6.3\times$ speedup and $3.3\times$ area efficiency, but only $1.4$--$1.6\times$ speedup in 5D/7D. Finally, enabling \textbf{Look-ahead Planning (LPE)} delivers the largest benefit, with the optimal depth varying by scene (see \Cref{tab:scene-params}): speedup reaches $15.0\times$ in 2D-City, $14.9\times$ in 2D-Maze, $14.1\times$ in 3D-Drone, and $3.9$--$4.2\times$ in 5D/7D-Arm. Area efficiency peaks at $7.8\times$ in 2D-City and is $2.0$--$2.2\times$ in 5D/7D. Power efficiency reaches $5.3\times$ in 2D-City and 2D-Maze and $1.4$--$1.5\times$ in 5D/7D. Across all workloads (geomean), the best look-ahead configuration achieves \textbf{$8.8\times$} speedup, \textbf{$4.6\times$} area efficiency, and \textbf{$3.1\times$} power efficiency over the ASIC w/ Tigris starting point. These results indicate that while each optimization contributes individually, they can all be integrated to achieve compounding PPA benefits.

% \begin{figure}[t]
%   \centerline{\includegraphics[width=\columnwidth]{figs/thread_benchmark.pdf}}
%   \vspace{-10pt}
%     \caption{\Name \hl{CPU Multi-thread Implementation Speedup} 
%     \textcolor{red}{to save space, add the CPU numbers to Table III } 
%     }
%     \label{fig:thread_benchmark}
%     \vspace{-15pt}
% \end{figure}

\begin{figure}[t]
\centerline{\includegraphics[width=\columnwidth]{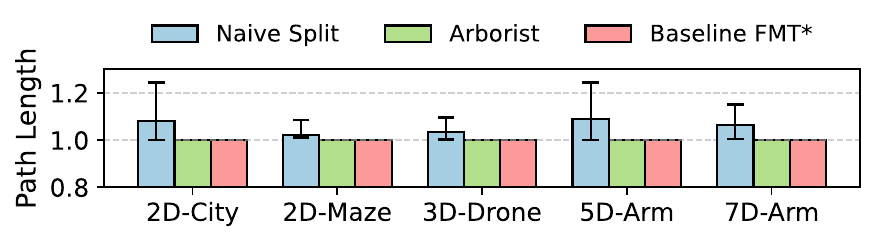}}
    \caption{Path Planning Quality.
    % Error bars indicating the 5th and 95th percentiles.
    %\jeff{@yaotian: last bar -> Baseline FMT*.}\yt{DONE}
    }
    \label{fig:path_quality}
\end{figure}

\noindent\textbf{Path Planning Quality.} 
%\jeff{@yaotian or 01: write a short para to expalin Fig. 16}
\Cref{fig:path_quality} compares the path planning
quality of \Name against the baseline FMT*, and a naive split
scheme without \Name Toolbox (or global competition management). While the naive split scheme also exploits parallel processing to reduce latency, it significantly degrades path quality by increasing the path length, on all evaluated maps. In contrast, \Name design (both in algorithm and on hardware) preserves the same optimal path cost as the original FMT*, and therefore achieves \textit{identical}
path quality while enabling high speedup, power and area efficiency. This requirement is strict in practice, because path quality
directly determines the operating time and energy consumption required for the whole robot system, rather than the motion planning
computation alone.

%\subsection{Components Results and Analysis}
%\jeff{For jeff to double check the following parameters below, need referring to each of them in Sec. 3 and Sec.4}

\subsection{Sensitivity Study}

\begin{figure}[t]
  \centerline{\includegraphics[width=\columnwidth]{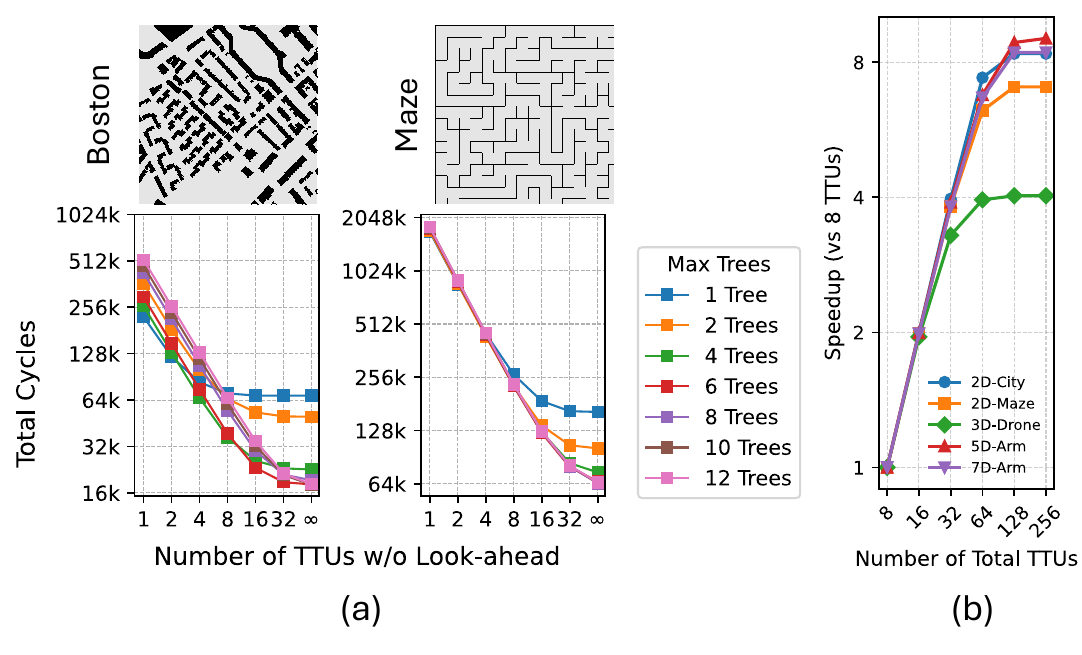}}
    \caption{(a) Effect of \texttt{MAX\_TREE}; (b) Number of TTUs on System Speedup.}
    \label{fig:max_tree}
\end{figure}

\noindent\textbf{Maximum Number of Trees (\texttt{MAX\_TREE}).}
\Cref{fig:max_tree}(a) sweeps \texttt{MAX\_TREE} and TTU count on Boston city and a Maze map~\cite{sturtevant2012benchmarks}. Both saturate beyond 6--8 trees, since the exploitable parallelism is ultimately bounded by obstacle structure. Boston (frequent wavefront reconvergence) incurs competition-management overhead when TTUs are scarce, but benefits substantially once enough TTUs are provisioned; Maze (diverging wavefronts) shows negligible overhead throughout. Higher-dimensional workloads follow the Boston trend, so picking \texttt{MAX\_TREE} according to obstacle density (typically 4–8) balances competition overhead against parallelism.

\noindent\textbf{Effect of TTU Count on System Speedup.}
\Cref{fig:max_tree}(b) reports end-to-end speedup vs.\ total TTU count with look-ahead planning enabled to exploit intra-tree parallelism. Compared to inter-tree parallelism alone (which saturates around 16--32 TTUs in (a)), look-ahead pushes near-saturation out to roughly 128 total TTUs. However, increasing the count from 64 to 128 yields only $2.1$--$30.9\%$ additional speedup across these workloads despite doubling the TTUs. The additional traversal and buffering resources increase area and power cost, motivating our choice of 64 TTUs as the default configuration.

\begin{figure}[t]
  \centerline{\includegraphics[width=\columnwidth]{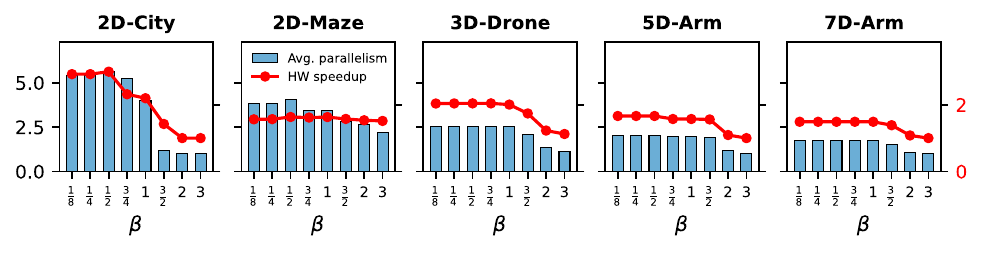}}
    \caption{$\beta$ sensitivity for the LZR Split Detector.}
    \label{fig:beta_sensitivity}
\end{figure}

\noindent\textbf{$\beta$ sensitivity for LZR Split Detector.}
\Cref{fig:beta_sensitivity} shows average parallelism (bars) and the hardware speedup (line) versus the split threshold~$\beta$. Lower thresholds split more aggressively, yielding higher parallelism and speedup, though speedup grows sub-linearly due to memory-bandwidth contention on Point Memory and issue-width limits of the Hierarchical Query Distributor. Across workloads, parallelism reaches $1.75$--$5.6\times$ and speedup $1.5$--$3.0\times$, with 2D-City peaking at $\beta{=}0.5$. The best operating point lies where the speedup curve flattens: $ \beta \in
[0.5,1.0]$ captures most of the achievable gain across all workloads.

% \begin{figure}[t]
%   \centerline{\includegraphics[width=\columnwidth]{figs/pca_split.pdf}}
%     \caption{\hl{PCA for Split Axis Selection}.}
%     \label{fig:pca_split}
% \end{figure}

% \begin{hlblock}
% \noindent\textbf{PCA for Split Axis Selection.}
% \Cref{fig:pca_split} shows the per-axis variance distribution (top) and gap coverage (defined in \Cref{eq:gap_cov}) as a function of $K$ (bottom) across five scenes. For low-dimensional scenes (e.g., 2D-City, 2D-Maze), variance is spread across all axes and all must be checked. The benefit emerges in higher dimensions: for 5D-Arm, the proximal joints $\theta_1$ and $\theta_2$ account for only 24\% of total variance but carry 98.3\% of the total gap coverage at $K=2$, because near-base obstacles create large forbidden regions in these two axes while distal joints ($\theta_3$--$\theta_5$) remain nearly unconstrained. The effect is more pronounced for 7D-Arm, where $K=2$ still captures 97.8\% of gap coverage out of seven axes. This means the split detector can skip 3 of 5 axes (5D) or 5 of 7 axes (7D) with less than 3\% gap coverage loss, reducing per-axis histogram overhead by 60--71\%.
% \end{hlblock}

\begin{figure}[t]
  \centerline{\includegraphics[width=1\columnwidth]{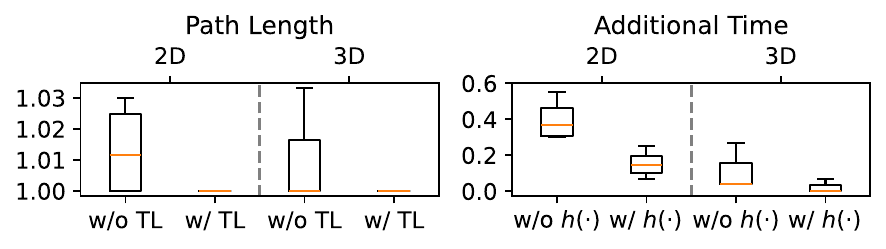}}
    \caption{Termination Logic (TL) ablation. $h(\cdot)$ denotes the heuristic used for cost estimation; one 3D no-TL point is 3.3\% above optimal.
    %~\jeff{@yaotian: one path cost with 2D (w/o TL, w/TL), 3D 2D (w/o TL, w/TL); one addtional time with 2D and 3D}\yt{DONE}
    %\jeff{@yaotian, is "path cost" the same as "path length"? If so please change to be consistent with other figs}\yt{DONE}
    }
    \label{fig:termination_ablation}
\end{figure}

\noindent\textbf{Effectiveness of Cross-tree Optimality Check.} As shown in \Cref{fig:termination_ablation}, omitting the proposed termination logic (TL) in \Name leads to up to 3.3\% additional path cost. In contrast, with TL enabled, all cases achieve the optimal path, demonstrating that our termination logic guarantees path quality. We also evaluate the benefit of incorporating a heuristic function $h(\cdot)$ for pruning, as described in \Cref{subsec:cross-tree-check} and illustrated in the right-hand figure. Using a simple $\ell_2$-norm heuristic as a lower bound on path cost reduces additional runtime by 45.5\% in the worst case and by 35.2\% on average, showing that our termination logic not only preserves optimality but also maximizes efficiency.

\label{subsubsec:exp-cross-tree-check}

% \begin{figure}[H]
%   \centerline{\includegraphics[width=0.8\columnwidth]{figs/ppa_overhead.pdf}}
%     \caption{}
% \end{figure}

\begin{figure}[t]
  \centerline{\includegraphics[width=\columnwidth]{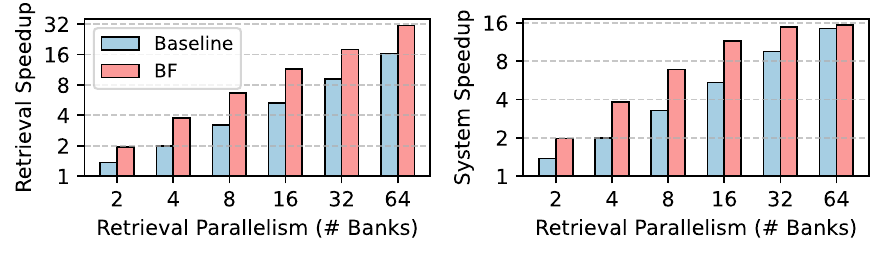}}
    \caption{Parallel Query Retrieval Speedup.}
    \label{fig:proposed_speedup_overhead}
\end{figure}

\noindent\textbf{Balance FIFO.}
\label{subsubsec:rf_vs_os}
The detailed results are shown in \Cref{fig:proposed_speedup_overhead}. Balance FIFO (BF) increases retrieval throughput by 90.2\% on average and by 116.7\% with $16$ banks.
When integrated into the full system (\Cref{fig:proposed_speedup_overhead}), higher retrieval speeds gradually shift the bottleneck away from memory, leading to diminishing overall gains. Nonetheless, a 16-bank Balance FIFO achieves a $2.12\times$ system speedup over the baseline.

% \begin{figure}[t]
%      \centerline{\includegraphics[width=1\columnwidth]{figs/comb_hierarchical_query_dist_para_bucket.pdf}}
%     \caption{(a) PPA trade-offs under different Hierarchical Query Distributor settings. %~\jeff{@yaotian, can we put fig. 18 and fig.19 side by side with a columwidth?} \yt{Done}
%     (b) Parallel Bucket Cache retrieval enables faster access and smaller cache size.
%     }
%     \label{fig:hier_query_dist_exp}
% \end{figure}

\begin{figure}[t]
     \centerline{\includegraphics[width=1\columnwidth]{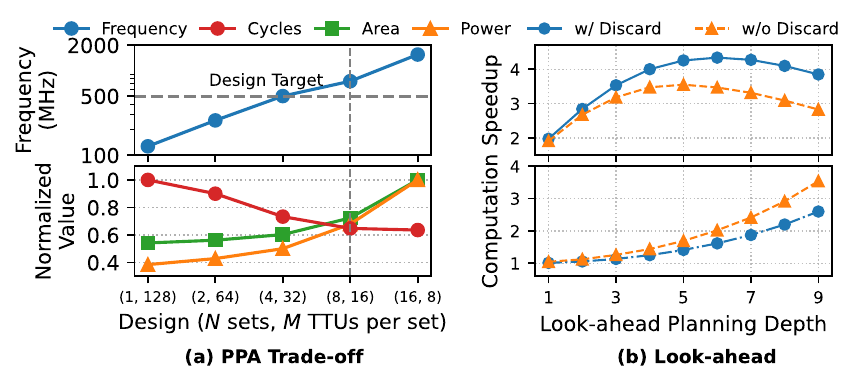}}
    \caption{(a) PPA trade-offs under different Hierarchical Query Distributor settings. 
    (b) Look-ahead Planning, where ``discard'' is whether to incorporate the Query Filter to discard repeated queries.
    }
    \label{fig:comb_ppa_lookahead}
\end{figure}

\noindent\textbf{Hierarchical Query Distributor.}
\Cref{fig:comb_ppa_lookahead}(a) sweeps $(N,M)$ configurations with 128 total TTUs and a 64-entry Query Filter ($N$ sets of $M$ TTUs, with a $64/N$-entry cache per set). The blue frequency curve is compared with the gray \SI{500}{\mega\hertz} target line. Larger $N$ eases timing and issues more parallel queries, but splitting the cache raises area/power with diminishing returns. The highlighted $(8,16)$ point is $35.3\%$ faster than $N{=}1$ and only $1.8\%$ slower than $N{=}16$ at acceptable overhead.

%\subsection{Parallel Bucket Cache Retrieval}

% \begin{figure}[htbp]
% \centerline{\includegraphics[width=0.6\columnwidth]{figs/bucket_cache_exp.pdf}}
% \caption{Parallel Bucket Cache retrieval enables faster access and smaller cache size.}
% \label{fig:bucket_cache_exp}
% \end{figure}

%As discussed in \cref{subsec:bucket_cache_para_update}, 

% Parallel query retrieval can also be leveraged for bucket cache misses. We evaluate this in \Cref{fig:hier_query_dist_exp} (b) by varying the retrieval speed, defined as the number of points fetched per cycle. 
% The results show that increasing cache size consistently reduces the total number of cycles, but there exists an effective saturation point: the first access will always incur a miss. With the original sequential scheme (1 point/cycle), the cache requires 32 buckets to achieve near-optimal performance. In contrast, with a parallel retrieval rate of 8 points/cycle, the same performance can be achieved with only 4 buckets, representing a 75\% reduction in cache size.

% This demonstrates that parallel retrieval not only improves performance but also reduces cache storage requirements, delivering both speed and area efficiency.

\noindent\textbf{Impact of Look-ahead Planning and Discard Mechanism.}
\Cref{fig:comb_ppa_lookahead} (b) shows the system speedup under different look-ahead depth $L$. 
With $L=1$, we achieve nearly $2\times$ speedup with almost no additional computation, since looking one step ahead is highly likely to succeed. As $L$ increases, speedup continues to grow but not linearly, because of the increase of false speculation causing redundant computation.

We also evaluate the effectiveness of the \textit{discard} mechanism provided by the Query Filter (\Cref{subsubsec:query_distributor}). As shown in the figure, discard substantially reduces redundant computation, allowing the system to achieve consistently higher speedup at the same $L$. 
%In particular, the gap widens at higher speculation levels, where redundant work would otherwise dominate.

\subsection{Programmability and Extensibility}

\begin{table}[t]
  \centering
  \caption{Latency breakdown (ms) for FMT*-family algorithms on CPU and \Accel, using 2D-City with 10k samples. Speedups use unrounded timings.}
  \label{tab:cpu_hw_breakdown_boston2d}
  \small
  \resizebox{\columnwidth}{!}{%
  \begin{tabular}{lccc!{\color{lightgray}\vrule}cccc}
  \toprule
   & \multicolumn{3}{c!{\color{lightgray}\vrule}}{CPU}
   & \multicolumn{4}{c}{\Accel} \\
  \textbf{Planner} & Build & Search & Total & Build & Search & Total & \textbf{Speedup} \\
  \midrule
  \textbf{FMT*}    & 3.04 & 119.96  & 123.0  & 0.373 & 0.052 & 0.425 & 289.6$\times$ \\
  \textbf{BFMT*}   & 3.11 & 46.22   & 49.3   & 0.373 & 0.016 & 0.389 & 126.7$\times$ \\
  \textbf{IAFMT*}  & 9.34 & 1555.66 & 1565.0 & 1.800 & 1.525 & 3.325 & 470.7$\times$ \\
  \textbf{IABFMT*} & 10.00 & 1376.00 & 1386.0 & 1.800 & 1.377 & 3.178 & 436.2$\times$ \\
  \bottomrule
  \end{tabular}%
  }
\end{table}

To evaluate \Accel's programmability, we replay four representative FMT*-family planners \textbf{FMT*}, \textbf{BFMT*}, \textbf{IAFMT*}, and \textbf{IABFMT*} on the same datapath with KD-tree, PQR, and LPE modules, and reconfigure only the control layer described in \Cref{subsec:programmability} (\Cref{fig:extensibility}). 
\Cref{tab:cpu_hw_breakdown_boston2d} reports the per-stage latency breakdown on 2D-City with 10k samples.

Across all four planners, \Accel delivers end-to-end
speedups ranging from $126.7\times$ (BFMT*) to $470.7\times$ (IAFMT*), with a $294.6\times$ geometric mean over the CPU baseline. The on-chip KD-tree constructor accelerates tree building by up to $8\times$, which is critical for the iterative variants (IAFMT*/IABFMT*) that must repeatedly rebuild the tree under anytime refinement and would otherwise be dominated by setup cost. 
To isolate the constructor's end-to-end benefit, we replace accelerator construction with CPU-side build while keeping the neighbor search on \Accel, yielding $7.28\times$/$8.03\times$/$3.27\times$/$3.58\times$ speedups for FMT*/BFMT*/IAFMT*/IABFMT* respectively, confirming that on-chip building is essential under dynamic replanning where the tree must be rebuilt frequently. The shared neighbor-search backend accelerates the CPU search stage by approximately three orders of magnitude across all planners, confirming that the PQR/LPE/KD-tree kernel remains effective regardless of the planner-specific control policy. These results show that \Accel is not hard-wired to the \Name algorithm: the same silicon serves as a configurable substrate for the broader FMT*-family at uniformly high efficiency.

\section{Related Work}

\noindent\textbf{Motion Planning Accelerators.}
Prior accelerators target motion planning end-to-end or its key subroutines~\cite{murray2016microarchitecture,murray2019programmable,shah2023energy,lian2018dadu,xu2019tigris,pinkham_quicknn_2020,exenberger2025caravan, hao2023blitzcrank, yang2020dadu}. Early systems target PRM~\cite{murray2016microarchitecture,PRM}; MOPED~\cite{huang2024moped} co-designs algorithm and hardware for RRT*~\cite{karaman2011sampling} to reduce both collision-check frequency and cost; RACOD~\cite{bakhshalipour2022racod} pairs a collision-detection accelerator with run-ahead search for grid-based A*~\cite{hart1968formal}; and MPAccel~\cite{shah2023energy} eliminates redundant collision checks via spatially aware early-exit scheduling.

%Prior work has proposed specialized hardware to accelerate motion planning either end-to-end or by offloading dominant kernels such as collision checking and nearest-neighbor queries~\cite{murray2016microarchitecture,murray2019programmable,shah2023energy,lian2018dadu,xu2019tigris,pinkham_quicknn_2020,exenberger2025caravan, hao2023blitzcrank, yang2020dadu}. 
%Early motion-planning accelerators targeted roadmap-style planning~\cite{murray2016microarchitecture,PRM}, while later systems co-designed hardware and algorithms for sampling-based planners (e.g., RRT*)~\cite{huang2024moped,karaman2011sampling} by restructuring the pipeline to reduce both the frequency and the cost of collision checking. 
%More recent designs further couple collision-detection acceleration with speculative/run-ahead search to improve latency in search-based planning~\cite{bakhshalipour2022racod,huang2024moped}.

\noindent\textbf{Wavefront-Style Planning (FMT*).}
FMT*~\cite{janson2015fastmarchingtreefast} expands an implicit random geometric graph as a cost-to-come ``wavefront'' via lazy dynamic-programming recursion, reducing collision checks and shifting the bottleneck toward neighbor search~\cite{janson2015fastmarchingtreefast,Sethian1996AFM}. A broad family of derivatives extends this principle~\cite{BFMT,GMT,xu2020informed,wang2023informed,allen2015toward,sucan2012the-open-motion-planning-library}, underscoring FMT*'s role as a foundational sampling-based planner and motivating accelerators that explicitly target wavefront expansion and high-rate nearest-neighbor queries.

\noindent\textbf{KD-tree Acceleration for Point Cloud Nearest-Neighbor Search.}
A complementary line of work accelerates KD-tree~\cite{bentley1999programming} nearest-neighbor (NN) search, a core primitive in 3D point-cloud processing where NN queries dominate runtime~\cite{xu2019tigris,pinkham_quicknn_2020,exenberger2025caravan, feng2022crescent}. These efforts exploit parallelism across queries and along the traversal, with memory-aware data layouts to mitigate bandwidth and locality bottlenecks~\cite{xu2019tigris,pinkham_quicknn_2020,exenberger2025caravan}; GPU-oriented techniques additionally support fast KD-tree (re)construction for dynamic point sets~\cite{zhou2008real}.

\section{Conclusion}
This work presents \Name, an algorithm–hardware co-design framework that exposes both inter- and intra-tree parallelism in the traditionally sequential FMT* algorithm via parallel multi-tree expansion, competition management, and look-ahead planning. Synthesized in 16 nm technology node and operating at 500 MHz, \Accel sustains more than $10^{3}\times$ speedup over CPU and gives geomean improvements of $8.8\times$ speedup, $4.6\times$ area efficiency, and $3.1\times$ power efficiency over an FMT* ASIC baseline. The corresponding factors relative to a state-of-the-art RRT*-based
MOPED accelerator are $8.6\times$/$1.0\times$/$1.8\times$, at comparable path cost.

\section*{Acknowledgments}
We sincerely thank the anonymous reviewers for their insightful feedback, which greatly improved the quality of this paper. This work was supported in part by the National Science Foundation (NSF) under Grant No.~2403409, No.~2239945, and the ASU Center for Semiconductor Microelectronics (ACME). Any opinions, findings, or recommendations expressed in this material are those of the authors and do not necessarily reflect the views of the funding agencies.

\begingroup
\setlength{\emergencystretch}{3em}
\appendices
\section{Artifact Appendix}

%%%%%%%%%%%%%%%%%%%%%%%%%%%%%%%%%%%%%%%%%%%%%%%%%%%%%%%%%%%%%%%%%%%%%
\subsection{Abstract}

This appendix summarizes the artifact evaluation for \Name and \Accel.
The artifact contains the motion-planning implementations and workloads,
cycle-level simulator, analysis scripts, and reference outputs used in this
work.  Release v1.1.4 provides a Docker-based top-level workflow that builds
and runs the package, regenerates the selected figures and tables, validates
the outputs, and reports a single pass or failure.  Detailed instructions are
included in the archive.

%%%%%%%%%%%%%%%%%%%%%%%%%%%%%%%%%%%%%%%%%%%%%%%%%%%%%%%%%%%%%%%%%%%%%
\subsection{Artifact check-list (meta-information)}

{\small
\begin{itemize}
  \item {\bf Algorithm: } FMT*, \Name, BFMT*, IAFMT*, and IABFMT*.
  \item {\bf Program: } C++ planners and Python simulation/analysis scripts.
  \item {\bf Compilation: } Docker (recommended), or GNU Make, GCC/G++ with
        C++17 support, and Boost headers for a native run.
  \item {\bf Data set: } Packaged 2D-City, 2D-Maze, 3D-Drone, 5D-Arm, and
        7D-Arm workloads.
  \item {\bf Run-time environment: } x86-64 Linux with Docker; Python~3.10 or
        newer is required only for a native run.
  \item {\bf Hardware: } Four logical CPUs and 8\,GiB RAM are recommended; no
        accelerator is required.
  \item {\bf Metrics: } Path cost, latency, cycles, speedup, area, power, and
        energy.
  \item {\bf Output: } Eight PDF figures, 17 CSV tables, logs, and a summary
        report.
  \item {\bf Disk space required (approximately): } 5\,GiB.
  \item {\bf Preparation time (approximately): } Under ten minutes when the
        Docker image or native dependencies are available.
  \item {\bf Experiment time (approximately): } About five minutes for the
        default workflow.
  \item {\bf Publicly available?: } Yes, at
        \url{https://doi.org/10.5281/zenodo.22313402}.
  \item {\bf Code license: } MIT for author-created code; third-party notices
        are included.
  \item {\bf Archived (DOI): }
        \url{https://doi.org/10.5281/zenodo.22313402}.
\end{itemize}
}

%%%%%%%%%%%%%%%%%%%%%%%%%%%%%%%%%%%%%%%%%%%%%%%%%%%%%%%%%%%%%%%%%%%%%
\subsection{Description}

\subsubsection{How to access}

We released v1.1.4 with a Docker-based workflow: download it from
\url{https://doi.org/10.5281/zenodo.22313402} and run
\texttt{./run-docker.sh}.
The archive is self-contained; no Git checkout,
submodule, private account, or data download is required.

\subsubsection{Hardware dependencies}

The default workflow runs on a standard x86-64 Linux machine.  Four logical
CPUs, 8\,GiB RAM, and 5\,GiB free disk are recommended.  A GPU, FPGA, RTL
simulator, and commercial EDA tools are not required.

\subsubsection{Software dependencies}

The recommended workflow requires Docker on an x86-64 Linux host.  Its first
image build requires Internet access to retrieve the base image and packaged
dependencies; the evaluator container then runs without network access.  For
the optional native workflow, compatible versions are GCC/G++~11 or newer,
GNU Make~4 or newer, and Python~3.10 or newer, together with Boost Container
and Boost Math headers.  The native workflow verifies the complete plotting
dependency closure at exact pinned versions and creates \texttt{.venv-ae} when
the active Python environment does not match.

\subsubsection{Data sets}

All five workload inputs are included in the archive.  Their sources, licenses,
and checksums are listed in \texttt{artifact/DATA.md}.

\subsubsection{Hardware artifacts}

The \texttt{hardware/} directory provides 11 selected RTL files (eight
report-matched components and three shared dependencies) together with eight
sanitized component synthesis summaries.

%%%%%%%%%%%%%%%%%%%%%%%%%%%%%%%%%%%%%%%%%%%%%%%%%%%%%%%%%%%%%%%%%%%%%
\subsection{Installation}

After downloading the artifact, run:

\begin{verbatim}
sha256sum -c arborist-ae-1.1.4.tar.gz.sha256
tar -xzf arborist-ae-1.1.4.tar.gz
cd arborist-ae-1.1.4
./run-docker.sh
\end{verbatim}

The Docker wrapper checks the package, builds the image, runs the experiments
in an isolated container, and validates the generated results.  The native
\texttt{./run.sh} workflow remains available for compatible host environments.
Manual installation and troubleshooting instructions are provided in
\texttt{artifact/INSTALL.md}.

%%%%%%%%%%%%%%%%%%%%%%%%%%%%%%%%%%%%%%%%%%%%%%%%%%%%%%%%%%%%%%%%%%%%%
\subsection{Suggested Experiment workflow}

The command \texttt{./run-docker.sh} is the recommended evaluator workflow.  It
runs the portable experiment set and regenerates the selected figures and
tables.  The key steps are:

\begin{itemize}
  \item run \texttt{./run-docker.sh} from the extracted archive root;
  \item confirm the final \texttt{AE PASS} and \texttt{21/21} stage summary;
  \item inspect \path{results/docker-ae/REPRODUCTION_REPORT.md}; and
  \item inspect the generated PDFs in \texttt{results/docker-ae/figures/} and
        CSVs in \texttt{results/docker-ae/tables/}.
\end{itemize}

The native workflow instead writes to \texttt{results/ae/}. The archive
names the eight experimental PDFs to match Figures~12--19 and
regenerates Table~III from the retained four-planner traces. Fixed/model
comparisons and separate sensitivity configurations remain identified in the
generated reports.

The optional paper-sample workflow is:

\begin{verbatim}
python3 artifact/reproduce.py \
  --mode full --jobs 24 \
  --output-dir results/full
\end{verbatim}

Commercial re-synthesis and fresh execution of external baselines are outside
this workflow.  Packaged fixed or modeled inputs remain identified in the
generated report.

%%%%%%%%%%%%%%%%%%%%%%%%%%%%%%%%%%%%%%%%%%%%%%%%%%%%%%%%%%%%%%%%%%%%%
\subsection{Evaluation and expected results}

A successful default run ends with:

\begin{verbatim}
AE PASS
Workflow stages: 21/21 passed
Generated figures: 8
Generated tables:  17
\end{verbatim}

This confirms that the released software builds and runs end-to-end, the five
FMT*/software-\Name workload pairs satisfy the registered path-cost check, and
the selected figures, tables, and machine-readable outputs pass validation.

%%%%%%%%%%%%%%%%%%%%%%%%%%%%%%%%%%%%%%%%%%%%%%%%%%%%%%%%%%%%%%%%%%%%%
\subsection{Experiment customization}

The map, sample count, tree count, split threshold, look-ahead depth, simulator
settings, and output directory can be changed.  See
\texttt{artifact/CUSTOMIZATION.md} and \texttt{artifact/REUSE.md} for examples.

%%%%%%%%%%%%%%%%%%%%%%%%%%%%%%%%%%%%%%%%%%%%%%%%%%%%%%%%%%%%%%%%%%%%%
\subsection{Methodology}

Submission, reviewing, and badging methodology:

\begin{itemize}
  \item \url{https://www.acm.org/publications/policies/artifact-review-and-badging-current}
  \item \url{https://cTuning.org/ae}
\end{itemize}
\endgroup

% Balance only the final bibliography page, not the appendix/reference transition.
\IEEEtriggercmd{\balance}
\IEEEtriggeratref{62}
\bibliographystyle{IEEEtran}
\bibliography{ref}

\end{document}